\documentclass[journal,twoside]{IEEEtran}
\usepackage{graphicx,hyperref,amsmath,color}
\newcommand{\etal}{{\it et~al.\/} }
\newcommand{\ie}{{\it i.e.\/} }
\newcommand{\eg}{{\it e.g.\/} }

\begin{document}
\title{Near-Field Radio Holography of Large Reflector
  Antennas}

\author{J.~W.~M.~Baars, 
        R.~Lucas, 
        J.~G.~Mangum, 
        and J.~A.~Lopez-Perez %
\thanks{Manuscript received February 21, 2006; revised November 30,
  2006.}%
\thanks{The performance results presented in this publication were part of a
comprehensive technical evaluation process used to evaluate the ALMA
prototype antennas which concluded in April 2005.}}

\markboth{IEEE Antennas and Propagation Magazine,~Vol.~49,
  No.~5,~October~2007}{Baars \MakeLowercase{\textit{et al.}}:
  Near-Field Radio Holography of Large Reflector Antennas}
%

\pubid{1045--9243/00\$00.00~\copyright~2007 IEEE}


\maketitle

\begin{abstract}
  We summarise the mathematical foundation of the holographic method
  of measuring the reflector profile of an antenna or radio
  telescope.  In particular, we treat the case, where the signal
  source is located at a finite distance from the antenna under test,
  necessitating the inclusion of the so-called Fresnel field terms in
  the radiation integrals. We assume a ``full phase'' system with
  reference receiver to provide the reference phase.  We describe in
  some detail the hardware and software implementation of the system
  used for the holographic measurement of the 12m ALMA prototype
  submillimeter antennas. We include a description of the
  practicalities of a measurement and surface setting.  The results
  for both the VertexRSI and AEC (Alcatel-EIE-Consortium) prototype
  ALMA antennas are presented. 
\end{abstract}


\IEEEpeerreviewmaketitle
%

\section{Introduction}
\label{intro}

\PARstart{L}{arge} reflector antennas, as those used in radio astronomy and
deep-space communication, generally are composed of a set of surface
panels, supported on three or more points by a support structure, often
called the backup structure. After assembly of the reflector it is
necessary to accurately locate the panels onto the prescribed
paraboloidal surface in order to obtain the maximum antenna gain. The
fact that some antennas have a "shaped"contour is irrelevant for the
purpose of our discussion. We are concerned with describing a method
which allows us to derive the position of the individual panels in
space and compute the necessary adjustments of their support points to
obtain a continuous surface of a certain prescribed shape.

The analysis by Ruze \cite{Ruze1966} of the influence of random errors in the
reflector contour on the antenna gain indicates that the RMS error
should be less than about one-sixteenth of the wavelength for
acceptable performance. Under the assumption that the errors are small
compared to a wavelength, randomly distributed with RMS value
\(\epsilon \), have a 
correlation length {\bfseries c} which is much larger than the
wavelength \(\lambda \), and much smaller than the reflector
diameter D, the relative decrease in aperture efficiency (or gain) can
be expressed by the simple formula 

\begin{equation}
\frac{\eta_A}{\eta_{A0}} = \exp\left\{-{\left(\frac{4\pi\epsilon}{\lambda}\right)}^2\right\},
\label{eq:etaa}
\end{equation}

\noindent{where} \({{\eta }_{A0}}\) is the aperture efficiency of the
perfect reflector. An error $\epsilon$ of \(\lambda \)/40 is required to limit
the gain loss to 10 percent; with an error of \(\lambda \)/16 the gain
is decreased to about half of the maximum achievable.

The setting of the reflector panels at accuracies better than 100
\(\mu \)m has required the development of measuring methods of
hitherto unsurpassed accuracy. It should be noted that these
measurements need to be done ``in the field'', which in the case of
millimeter radio telescopes generally means the hostile environment of
a high mountain site.  One versatile, and by now widely used method is
normally called ``radio holography''. The method makes use of a
well-known relationship in antenna theory: the far-field radiation
pattern of a reflector antenna is the Fourier Transformation of the
field distribution in the aperture plane of the antenna. Note that
this relationship applies to the amplitude/phase distributions, not to
the power pattern. Thus, if we can measure the radiation pattern, in
amplitude and phase, we can derive by Fourier Transformation the
amplitude and phase distribution in the antenna aperture plane with an
acceptable spatial resolution. Bennett \etal \cite{Bennett1976}
presented a sufficiently detailed analysis of this method to draw the
attention of radio astronomers. Scott \& Ryle \cite{Scott1977} used
the new Cambridge 5 km array to measure the shape of four of the eight
antennas, using a celestial radio point source and the remaining
antennas to provide the reference signal. Simulation algorithms were
developed by Rahmat-Samii \cite{RahmatSamii1985} and others, adding to
the practicability of the method.  Using the giant water vapour maser
at 22 GHz in Orion as a source Morris \etal \cite{Morris1988}
achieved a measurement accuracy of 30 $\mu$m and were able to set the
surface of the IRAM 30-m millimeter telescope to an accuracy of better
than 100 $\mu$m. 

\pubidadjcol 

Artificial satellites, radiating a beacon signal at a fixed frequency
have also been used as farfield (${R_f} = \frac{2 {D^2}}{\lambda}$)
signal sources. Extensive use has been made of synchronous
communication satellites in the 11 GHz band \cite{Godwin1986},
\cite{Rochblatt1992}. These transmitters of course do not
provide the range of elevation angles accessible with cosmic
sources. Some satellites, notably the LES (Lincoln Experimental
Satellite) 8 and 9, have been used for radio holography of millimeter
telescopes \cite{Baars1999}. They provided a signal at the
high frequency of 37 GHz and with their geo-synchronous orbit moved
over some 60 degrees in elevation angle. Unfortunately, both
satellites are no longer available. Lacking a sufficiently strong
source in the farfield, we have to take recourse to using an
earth-bound transmitter. In practice these will be located at a
distance of several hundreds of meters to a few kilometers and be at
an elevation angle of less than 10 degrees. Clearly, these are in the
nearfield of the antenna, requiring significant corrections to the
received signals.  In particular, the phase front of the incoming
waves will not be plane and it contains higher order terms in the
radial coordinate of the antenna aperture. These must be corrected
before the Fourier transformation can be applied. We treat these
corrections in detail in this paper.

Successful measurements on short ranges have been reported for the
University of Texas millimeter telescope \cite{Mayer1983}, the IRAM
30-m telecope \cite{Morris1988b}, the JCMT \cite{Hills2002} and the
ASTE antenna of NAOJ \cite{Ezawa2000}.

ALMA (Atacama Large Millimeter Array) is a new large aperture
synthesis array for submillimeter astronomy consisting of 50 high
accuracy antennas of 12 m diameter. The instrument is under
construction at 5000 m altitude in the Atacama desert of northern
Chile. ALMA is a collaboration of North America and Europe with
participation of Japan.  Two prototype antennas were procured and
erected at the site of the Very Large Array of NRAO in New Mexico. The
results of an extensive evaluation program of these antennas has been
presented by Mangum \etal \cite{Mangum2006}. The reflector surface
accuracy was specified at 20-25 $\mu$m, requiring a measurement method
with an accuracy of 10 $\mu$m or better. This was achieved with a
near-field holographic system using a transmitter at a wavelength of 3
mm and at a distance of only 315 m from the antennas at an elevation
angle of 9 degrees. Here we describe these measurements in some
detail.

\section{The Mathematics of Radio Holography}

\begin{figure}
\resizebox{\hsize}{!}{
\includegraphics{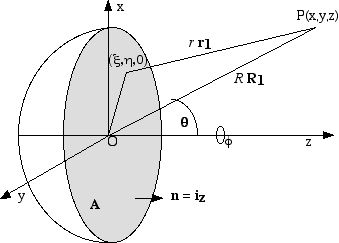}}
\caption{Geometry of the aperture integration method for finite
distance to the field point P.}
\label{fig:geofig}
\end{figure}

The reciprocity theorem describes the equivalency between the
characteristics of a transmitting and receiving antenna. Thus both
concepts will be used in the following treatment depending on the
specific aspect under description. We shall not repeat here the
fundamental analysis which leads from Maxwell's equations to the
``physical optics'' representation of the characteristics of the
reflector antenna (\eg\ \cite{Silver1949}, \cite{Rusch1970}). 
Our discussion below essentially follows the treatment by Silver
\cite{Silver1949}. The basic expression, linking the radiation 
function \(f(x,y,z) \) at a point P in space with the 
field distribution \(F(\xi,\eta )\) over the aperture plane of the
antenna, is written as (see Fig.~\ref{fig:geofig} for the geometry) 

\begin{eqnarray}
\lefteqn{f(x,y,z) = \int \frac{{F(\xi ,\eta ) e^{-i kr}}}{4\pi r} } \nonumber \\
     && \left[\left(i k+\frac{1}{r}\right) \mbox{\boldmath $i_{z}
    \cdot {r_1}$}+i k \mbox{\boldmath $i_{z} \cdot {s}$}\right] d \xi d \eta,
\label{eq:field1}
\end{eqnarray}
		
\noindent{where} the integration is extended over the aperture area,
\(k = 2 \pi /\lambda \) and the unit vectors are as indicated in
Fig.~\ref{fig:geofig} (with
{\bfseries s} the propagation vector of the wave field in the
aperture). This relation assumes that the aperture is large in units
of the wavelength. This general expression can be simplified depending 
on the distance of the field point P from the aperture plane. We
discern the so-called far-field region (Fraunhofer diffraction),
near-field region (Fresnel diffraction) and the evanescent-wave zone up
to a few wavelengths from the reflector. In the evanescent-wave zone,
which does not concern us here, no approximations are allowed. 

If the field point P is sufficiently far away from the aperture, the
following simplifications can be introduced in the evaluation of
Eq.~\ref{eq:field1}:

\begin{enumerate}
\item The term $\frac{1}{r}$ is ignored with respect to \(k\) in
	the bracketed term.
\item The term $\frac{1}{r}$ outside the brackets is replaced by the
  reciprocal distance $\frac{1}{R}$ from the aperture center to the
  field point P.  
\item The term \(\mbox{\boldmath ${i_z} \cdot {r_1}$}\) can be
	approximated by \(\mbox{\boldmath${i_z} \cdot {R_1}$} =
	\cos\theta \) with
	\(\mbox{\boldmath $R_1$}\) the unit vector from the origin to the
	field point.
\item the term \(\mbox{\boldmath ${i_z} \cdot s$}\) represents a deviation from
  uniform phase over the aperture. If these are small, this term can be
  assumed to be equal to one over the aperture.  
\end{enumerate}

\noindent{With} these approximations Eq.~\ref{eq:field1} is
simplified to

\begin{equation}
f(x,y,z) = \frac{i}{2 \lambda R} \int F(\xi ,\eta) \left[\cos \theta +1\right]
   {e^{i k r}} d \xi d \eta.
\label{eq:field2}
\end{equation}
			
\noindent{For} the distance r from any point in the aperture to the
field point P we have (see Fig.~\ref{fig:geofig}) 

\begin{equation}
r = {{\left\{{{(x - \xi )}^2}+{{(y-\eta )}^2} + {z^2}\right\}}^{0.5}}.
\label{eq:r1}
\end{equation}

\noindent{Writing} the coordinates of the field point P(x,y,z) in spherical
coordinates, we obtain

\begin{eqnarray}
x &=& R \sin \theta \cos\phi \equiv Ru, \nonumber \\
y &=& R \sin \theta \sin\phi \equiv Rv, \nonumber \\
z &=& R \cos \theta =R {\sqrt{(1-{u^2}-{v^2})}},
\label{eq:sphericalcoord}
\end{eqnarray}
	
\noindent{where} we have also introduced the {\bfseries direction
cosines} of the field point 

\begin{displaymath}
P(u,v)=(\sin\theta \cos\phi, \sin\theta \sin\phi).
\end{displaymath}

\noindent{Thus}, Eq.~\ref{eq:r1} can be written as

\begin{eqnarray}
r &=& {{\{{{(R u - \xi )}^2}+ {{(R v - \eta
	)}^2}+{R^2}(1-{u^2}-{v^2})\}}^{0.5}} \nonumber \\
  &=& R {{\left\{1-2 \frac{u \xi +v \eta }{R}+\frac{{{\xi }^2}+{{\eta
	}^2}}{{R^2}}\right\}}^{0.5}}.
\label{eq:r2}
\end{eqnarray}

\noindent{The} series expansion of Eq.~\ref{eq:r2} yields

\begin{eqnarray}
r &\approx& R - (u \xi + {v\eta })+\frac{{{\xi }^2}+{{\eta }^2}}{2
  R}-\frac{{{({{\xi }^2}+{{\eta }^2})}^2}}{8{R^3}}- \nonumber \\
  && \frac{{{(u \xi +v\eta )}^2}}{2 R}+ 
     \frac{({{\xi }^2}+{{\eta }^2}) (u \xi +v \eta)}{2{R^2}}-\ldots
\label{eq:r3}
\end{eqnarray}

\subsection{The Far-Field Approximation (Fraunhofer Region)}

In the {\bfseries far-field} situation, where R tends to
infinity and \(\mbox{\boldmath $R_1$}\) and \(\mbox{\boldmath $r_1$}\)
are essentially parallel, the variation of r in the exponent of
Eq.~\ref{eq:field2} can be reduced to the linear part of
Eq.~\ref{eq:r3} 

\begin{equation}
r = R - (u \xi + {v\eta })
\label{eq:r4}
\end{equation}
			
\noindent{Additionally}, considering that for a high gain antenna the
angular region of interest is confined to small values of $\theta$, we
can write Eq.~\ref{eq:field2} with \(\cos\theta = 1\), which is
valid to 0.1\% for angles up to 3 degrees off-axis.  The radiation
integral (Eq.~\ref{eq:field2}), then becomes

\begin{equation}
f(u,v) =\frac{i}{\lambda} \frac{{e^{-{ikR}}}}{R} \int F(\xi,\eta) 
\exp\{-i k (\xi u + \eta v)\} d\xi d\eta
\label{eq:fpuv}
\end{equation}

\noindent{where} the integration is performed over the aperture
A.  We see from Eq.~\ref{eq:fpuv} that there exists a Fourier
Transformation relationship between \(f(u,v)\) and \(F(\xi ,\eta )\).
Ignoring the term $\frac{i}{\lambda}$, the inverse Fourier
transformation can thus be written as

\begin{equation}
F(\xi ,\eta ) = \frac{1}{4\pi} \frac{{e^{ikR}}}{R} \int f(u,v) \exp\{i k (u\xi +v \eta )\} du dv,
\label{eq:fchieta}
\end{equation}

\noindent{where} the integration in principle has to be performed over
a closed surface, surrounding the aperture. Thus a knowledge of the entire
far-field pattern \(f(u,v)\) both in amplitude and in phase provides a
description of the complex field distribution \(F(\xi ,\eta )\)
over the aperture plane of the antenna, also in amplitude and phase.
This forms the basis of the so-called \textit{radio holographic
  measurement} of the shape of a reflector antenna.  Deviations from a
uniform phase function over the aperture are thereby linked to local
errors in the prescribed contour of the reflector surface.

It is interesting to note that Silver devotes a lengthy discussion to
this Fourier Transform relationship (\cite{Silver1949}, Ch. 6.3), but
concludes that the practical 
application is limited by the fact that the far-field pattern is only
prescribed in power. Thus the phase function of \(f(u,v)\) 
would be arbitrary and the aperture distribution cannot be
uniquely determined. It was Jennison \cite{Jennison1966} who mentioned
the same relation and its practical usefulness, pointing 
out that the amplitude and phase can both be measured with an
interferometer. When Silver wrote his text in the mid-1940s, radio
interferometry had not yet been developed. 

In most cases it will be impossible, or in any case impractical, to
measure the far-field pattern over the entire sphere. The Nyquist
sampling theorem shows however that a measurement of the pattern out
to an angle \(\Theta \) \(=\) n \({{\Theta }_A}\) from the beam axis
yields the aperture distribution with a spatial resolution of \(\delta
= \frac{D}{n}\), where \({{\Theta}_A} \approx \frac{\lambda}{D}\) is the 
half-power beam-width, D is the aperture diameter, and \(\lambda \) the
wavelength.

\subsection{The Near-Field Approximation (Fresnel Region)}
\label{sec:fresnel}

In the {\bfseries near-field} region, which corresponds to the Fresnel
region in optical diffraction (\eg\ \cite{Born1970}) most of the
simplifications leading to Eq.~\ref{eq:field2} can still be used,
as long as r is at least several aperture diameters large.
However, the variation in r over the aperture must now include
higher-order terms in Eq.~\ref{eq:r3} and be maintained in the
exponent (phase) term of the integral.  Normally, for the Fresnel
region analysis, the 
series is stopped after the quadratic term, which preserves the first
three terms of the series in Eq.~\ref{eq:r3}.  Thus the near-field
(Fresnel region) expression can be written in the form of the
following radiation integral, which is the well-known Fresnel
diffraction integral in two coordinates:

\begin{eqnarray}
\lefteqn{f(u,v) = \frac{i}{\lambda}\frac{{e^{i k R}}}{R}\int F(\xi,\eta ) } \nonumber \\
       && \exp\left\{i k \left[-(u \xi +v \eta )+\frac{{{\xi
      }^2}+{{\eta }^2}}{2R}\right]\right\} d\xi d \eta.
\label{eq:fuv}
\end{eqnarray}

In the application of holography in the near-field we want to derive
the complex aperture field distribution from the measured near-field
pattern. Thus the inverse Fourier Transformation of
Eq.~\ref{eq:field2} will be our point of departure, where
Eq.~\ref{eq:r2} is the expression for the finite distance from a point
in the aperture to the point where the signal source is located. Thus
we have the inverse of Eq.~\ref{eq:field2}

\begin{equation}
F(\xi,\eta ) = \frac{i}{\lambda	R} \int f(u,v) \exp(-i k r) du dv.
\label{eq:fchieta2}
\end{equation}

\noindent{where} R is the distance from the antenna aperture center to
the holography signal source.  We maintain all terms of
Eq.~\ref{eq:r3} in order to make an estimate of the error with respect
to the usual Fresnel approximation.

We rewrite Eq.~\ref{eq:r3} as follows:

\begin{equation}
r \approx R - (u\xi + v\eta ) + \delta p_1 (\xi,\eta) + \epsilon
\label{eq:r3again}
\end{equation}

\noindent{where} we define the terms, which are independent
of the integration variables, as the variable \({{{\delta p}}_1}\):

\begin{equation}
{\delta p_1}(\xi,\eta ) = \frac{\xi^2+\eta^2}{2 R}-\frac{(\xi ^2+\eta ^2)^2}{8 R^3},
\label{eq:deltap1}
\end{equation}

\noindent{while} the other terms in higher powers of \({(u,v)}\) are
collected under the variable \(\epsilon \). 

\begin{equation}
\epsilon =-\frac{{{(u\xi +v \eta )}^2}}
{2R}+\frac{\big({{\xi}^2}+{{\eta }^2}\big) (u \xi +v \eta ) }{2{R^2}},
\label{eq:epsilon}
\end{equation}

\noindent{Substitution} of Eq.~\ref{eq:r3} into Eq.~\ref{eq:fchieta2}
yields

\begin{eqnarray}
\lefteqn{F(\xi ,\eta ) = \frac{i}{\lambda} \frac{{e^{-i k R}}}{R} \exp\{-i k
{{\delta p}_1}(\xi ,\eta )\}} \nonumber \\
              && \int f(u,v) \exp\{i k (u \xi + {v\eta })\} {e^{-i k
    \epsilon}} du dv,
\label{eq:fchieta3}
\end{eqnarray}

\noindent{The} terms in $\epsilon$ ``modify'' the direct
Fourier transformation of Eq.~\ref{eq:fchieta3}.

\subsubsection{The $\mathbf \delta p_1$ Term}
\label{deltap1}

\begin{figure}
\resizebox{\hsize}{!}{
\includegraphics[trim=50 300 50 70]{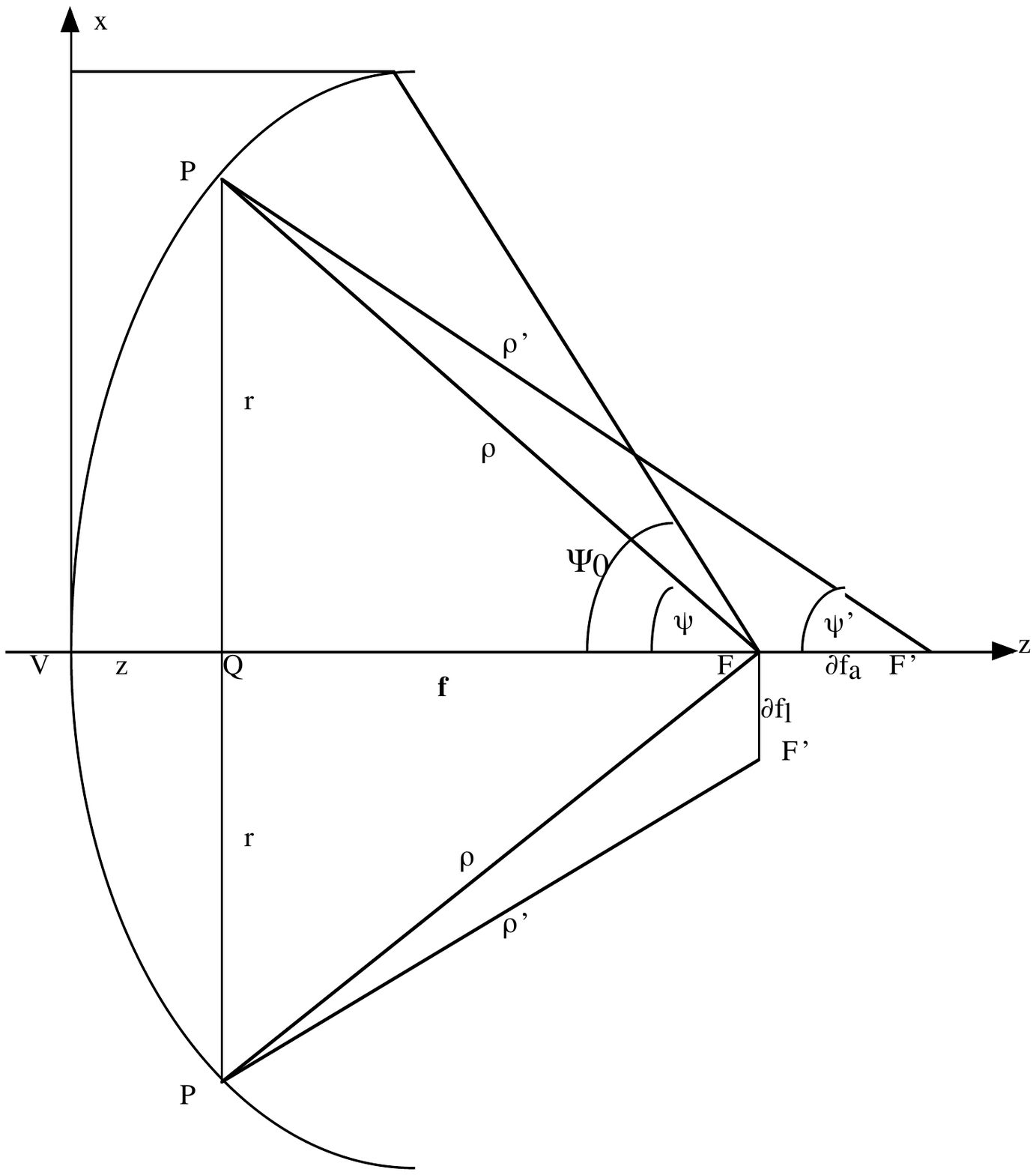}}
\caption{The geometry of axial (upper half) and lateral (lower half)
  displacement of the feed from the focus of a parabola.}
\label{fig:feeddisplacement}
\end{figure}

The first path-length term \({{{\delta p}}_1}\) causes a rapidly
varying phase variation over the aperture, which can be compensated to
a large degree by an axial displacement of the feed.  
In Fig.~\ref{fig:feeddisplacement} we illustrate the geometry of both a
lateral \({{\delta f}_l}\) and an axial \({{\delta f}_a}\)
displacement of the feed from the primary focus. We need to calculate
the difference in path-length (\({\rho^\prime - \rho}\)) as a function
of the radial position r of the point P.  The residual path-length
error at P with respect to the ray to the vertex V will then be given
by

\begin{eqnarray}
{\delta p_2}(\xi ,\eta ) &=& \delta p_2(r) \nonumber \\
                             &=& \left(\rho^\prime - \rho\right)_P -
                             \left(\rho^\prime - \rho\right)_V,
\label{eq:deltap2-1}
\end{eqnarray}

\noindent{where} \({r^2 = \xi^2 + \eta^2}\).  Applying Pythagoras' law
to the upper triangle PQF$^\prime$ and using the defining relation for
the parabola \({r^2 = 4fz}\), where \({z = VQ}\), we find for the
path-length variation due to an axial defocus $\delta f$ away from the
reflector

\begin{eqnarray}
{{{\delta p}}_2}(\xi ,\eta ) &=& {{\left\{{{\xi}^2}+{{\eta }^2}+{{\left(f-
	\frac{{{\xi }^2}+{{\eta }^2}}{4 f}+{\delta
	  f}\right)}^2}\right\}}^{0.5}} - \nonumber \\
&& \left\{f+\frac{{{\xi
    }^2}+{{\eta}^2}}{4 f}+{\delta f}\right\}.
\label{eq:deltap2}
\end{eqnarray}

\begin{figure}
\resizebox{\hsize}{!}{
\includegraphics{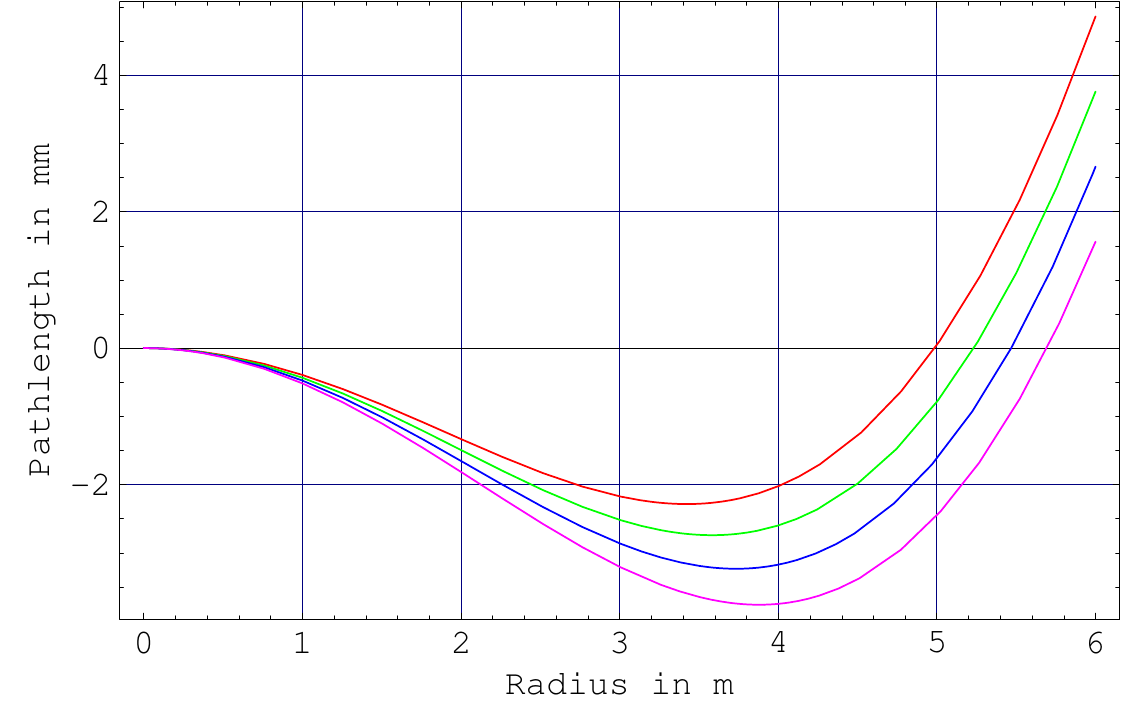}}
\caption{The residual path-length error (\({\delta p_1}+{\delta
    p_2}\)) in mm for a distance R \(=\)
315 m to the holography transmitter and the ALMA 12-m diameter antenna
with \(\frac{f}{D} = 0.4\). The parameter is the axial defocus \(\delta \)f
\(=\) 96, step 2, 102 mm, from top to bottom.}
\label{fig:respath}
\end{figure}

We want to minimise the sum of the two terms (\({\delta p_1}+{\delta
  p_2}\)) (Eqs.~\ref{eq:deltap1} and \ref{eq:deltap2}) by
choosing the appropriate value of \(\delta f \). Because of the (\(\xi
\),\(\eta \))-dependence there will be a residual
path-length error, which we must apply to the result of the Fourier
Transformation.  Fig.~\ref{fig:respath} shows the residual
path-length error (\({\delta p_1}+{\delta p_2}\))
for several choices of \({\delta f}\) for the
geometry of the ALMA antennas and the actual distance to the
holography transmitter.  A value of 96--98 mm limits the error to $\pm
3$ mm over the aperture for R \(=\) 315 m. This remaining error must
be introduced in the mathematical analysis of the data according to
the curve shown in Fig.~\ref{fig:respath}.  This is a correction to
the aperture phase distribution, obtained after the Fourier
Transformation of the measured beam pattern.

\subsubsection{The $\mathbf \epsilon$ Term}
\label{epsilon}

The higher order terms in Eq.~\ref{eq:epsilon} ($\epsilon$),
containing the integration variables (\(u,v)\), constitute a small
path-length error which 
adds a phase term to the integral in Eq.~\ref{eq:fchieta3} of the
form  

\begin{eqnarray}
\exp(- i k \epsilon ) &\approx& 1- i k \epsilon \nonumber \\
&=& 1 - i k\left\{u
\frac{\xi ({{\xi }^2}+{{\eta }^2})}{2{R^2}}+v\frac{\eta ({{\xi
    }^2}+{{\eta }^2})}{2 {R^2}}- \right. \nonumber \\
&& \left. {u^2}\frac{{{\xi }^2}}{2
  R}-{v^2}\frac{{{\eta }^2}}{2 R}- u v \frac{\xi \eta }{R}\right\}.
\label{eq:phaseterm}
\end{eqnarray}
		
\begin{figure}
\resizebox{\hsize}{!}{
\includegraphics{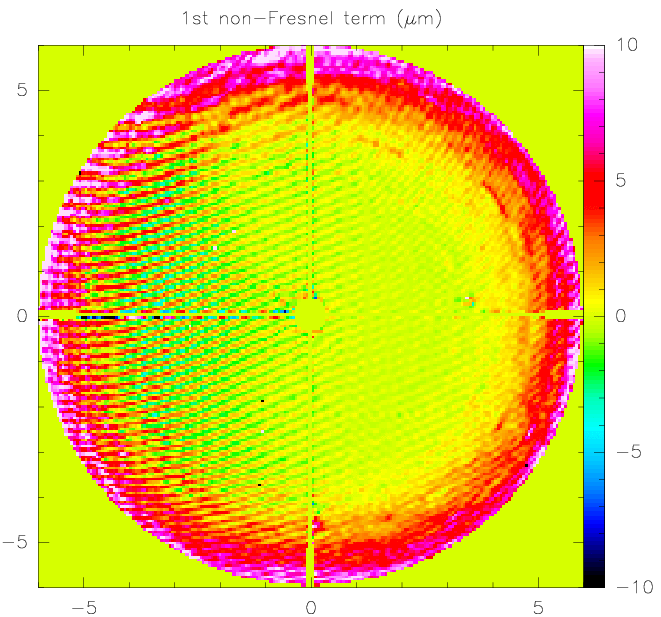}}
\caption{Non-Fresnel correction terms over the aperture of the 12 m
  diameter ALMA antenna. Horizontal and vertical axes units are meters
  with color showing surface error in $\mu$m shown at right.  The departures
  from circular symmetry are due to the effects of the actual surface
  errors present in the map.}
\label{fig:nonfresnel}
\end{figure}

\noindent{It} is seen that this correction involves the calculation of
five additional integrals, which look like Fourier transformations, but
aren't really {\itshape bona fide} Fourier transformations. When all
the integrals of 
Eq.~\ref{eq:phaseterm} are evaluated, it turns out that the
contribution of $\epsilon$ amounts to 2$\mu$m of path-length over most
of the aperture, reaching a peak value of 10$\mu$m at the very edge of
the aperture. This is illustrated in Fig.~\ref{fig:nonfresnel}. In a high
accuracy measurement, where the aim is to achieve a measurement accuracy
of better than 10 $\mu$m, it is advisable to include this correction term. 

\subsubsection{Dependence on Aperture Plane Reference}
\label{aperref}

\begin{figure}
\resizebox{\hsize}{!}{
\includegraphics[scale=0.84, angle=-90]{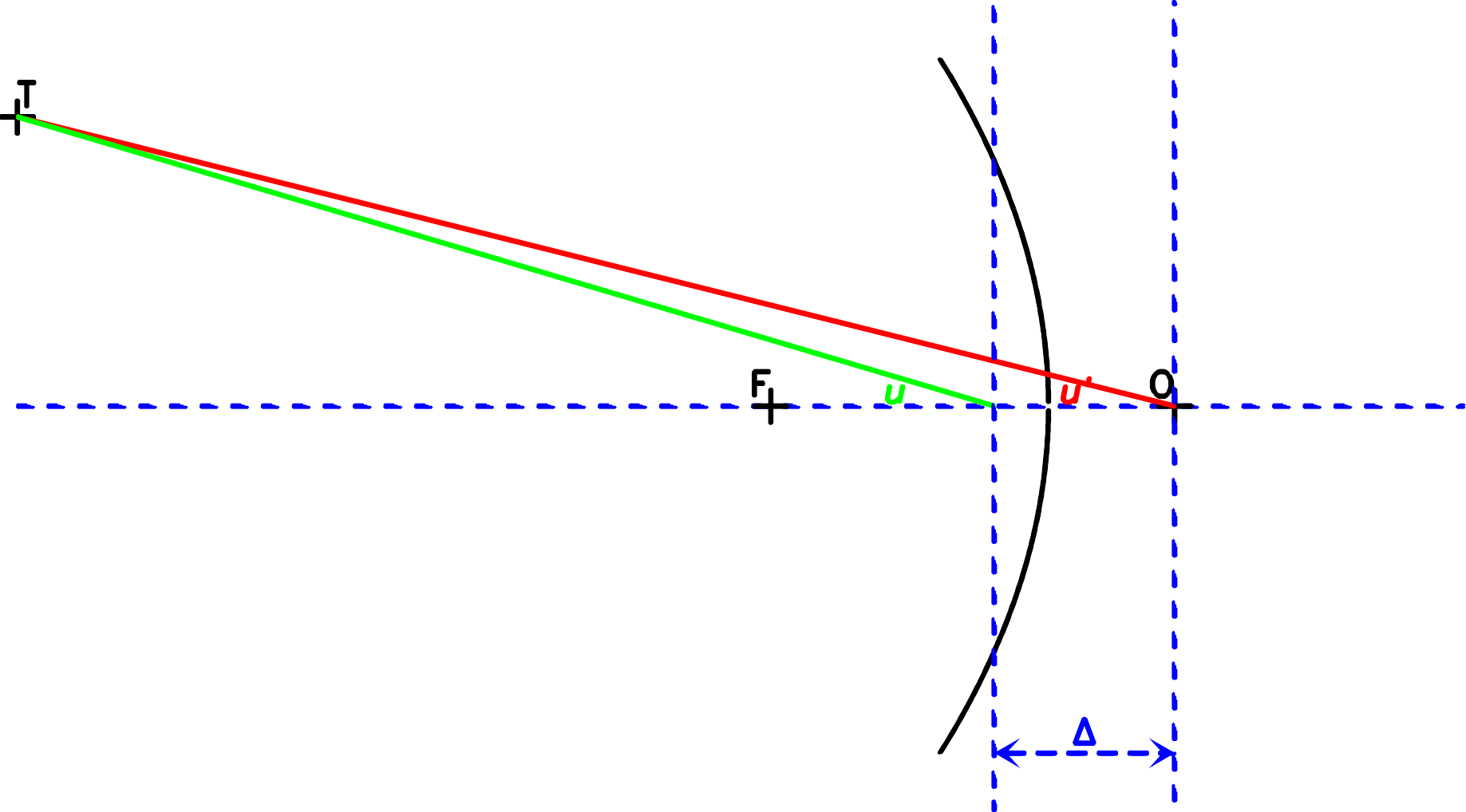}}
\caption{Illustration of the geometry of selected aperture plane and
  antenna rotation axis. T is the position of the transmitter while
  scanning the antenna.} 
\label{fig:parallax}
\end{figure}

In the corrections for the finite distance to the transmitter (the
near-field corrections), we define the aperture plane at a convenient
location, normally halfway between the vertex and the edge of the
reflector. We have taken the center of this aperture plane as the
origin of the coordinate system. In most antennas there is a
significant distance between this plane and the axes of rotation for
the movement of the reflector (see Fig.~\ref{fig:parallax}).
From this figure we see that there is a ``parallax'' effect between
the adopted direction cosines (u, v) and those given by the antenna
scanning coordinates (u$^\prime$, v$^\prime$), given by the relations

\begin{eqnarray}
u &=& u^\prime \left(1 + \frac{\Delta}{R}\right) \nonumber \\
v &=& v^\prime \left(1 + \frac{\Delta}{R}\right),
\label{fig:uv}
\end{eqnarray}
 
\begin{figure}
\resizebox{\hsize}{!}{
\includegraphics{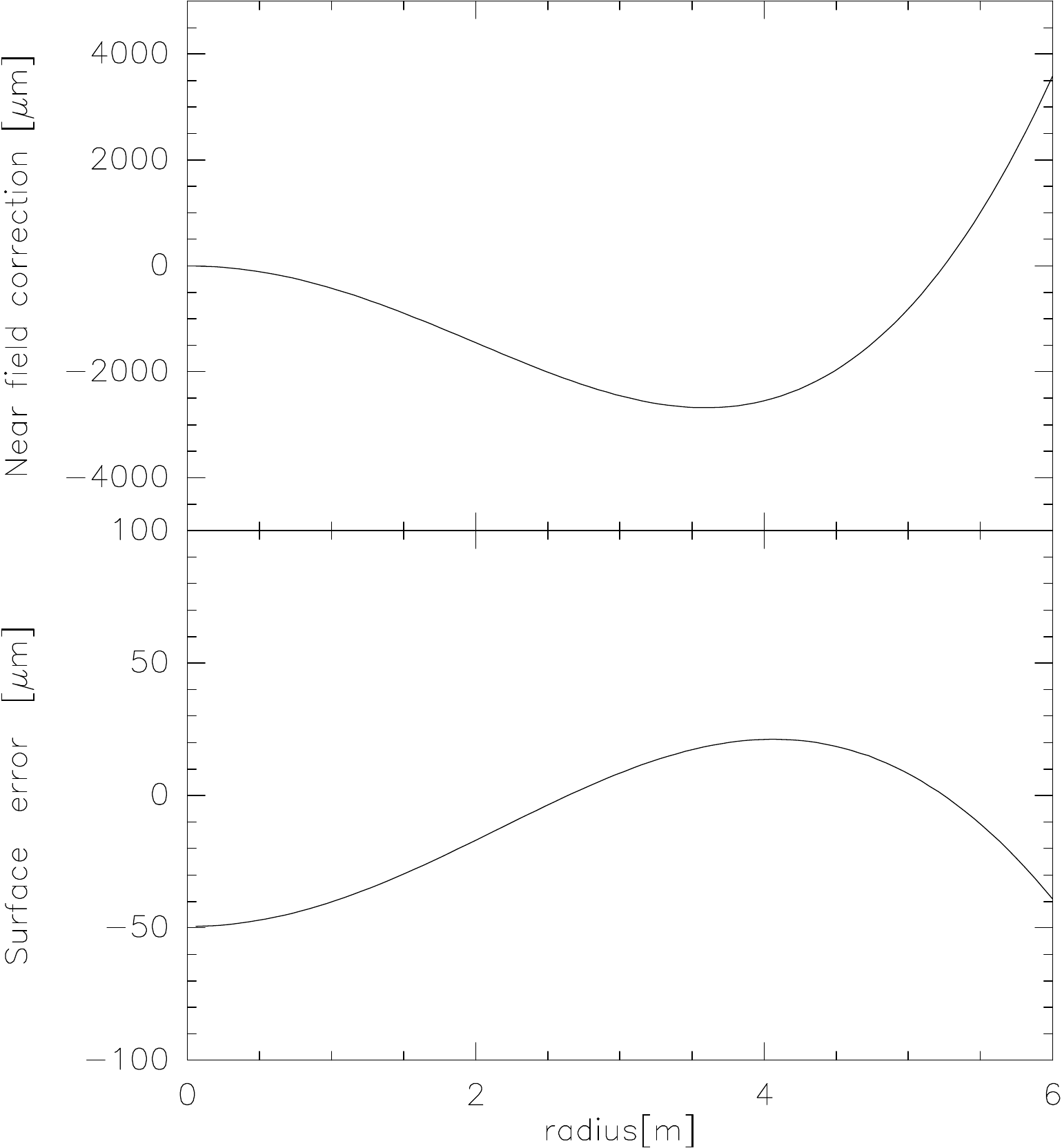}}
\caption{Top: The near-field path-length correction as a function of
  radius assuming $\Delta $ = 3.1 m and R = 315 m.  Bottom: The
  surface error resulting from the misregistration of the radial
  coordinate (\ie\ ignoring the difference between direction
  cosines \(u\) and \(u^\prime\)).}
\label{fig:geodistance}
\end{figure}

\noindent{where} $\Delta$ is the distance between rotation axis and
aperture plane. We use the scanning coordinates (u$^\prime$,
v$^\prime$), read from the antenna encoders, to calculate the position
of the points in the aperture plane. If (u, v) in the Fourier integral
(Eq.~\ref{eq:fchieta3}) are approximated by
\((u^\prime,v^\prime)\), the scale of the aperture map will be
overestimated by the factor \({\left(1 +
  \frac{\Delta}{R}\right)}\). The result of 
this is that the near-field correction for each pixel in the map is not
evaluated at the correct radius. This causes a path-length error
proportional to the derivative of the near-field correction with
respect to the radial coordinate. Fig.~\ref{fig:geodistance}
illustrates the magnitude of this effect for the case of our geometry,
where $\Delta\approx 3.1$m, \ie\ about 1\% of the distance R to the
transmitter. The error is significant, causing a surface error as
a function of radius as shown in the lower part of
Fig.~\ref{fig:geodistance}; its RMS value is 18$\mu$m in our case,
significant with respect to the required setting accuracy.  Such a
donut-shaped systematic deviation was indeed found in our original
maps, once we applied the geometry correctly. It was of course treated
properly in the final measurements and surface setting.

\subsubsection{Dependence on Focal Deviation}
\label{focdev}

It is possible that during the measurement the receiver feed is not
located in the optimum focal position. With reference to
Fig.~\ref{fig:feeddisplacement} and Eq.~\ref{eq:deltap2} it can
be shown that the path-length error caused by an axial defocus of
\(\delta\)z is given by

\begin{equation}
{{{\delta p}}_z} = {\delta z} \left\{1 - \frac{1-\frac{{{\xi
      }^2}+{{\eta }^2}}{4 {f^2}}+\frac{{\delta
      f}}{f}}{{\sqrt{\frac{{{\xi }^2}+{{\eta }^2}}{4
	{f^2}}+{{\big(1-\frac{{{\xi }^2}+{{\eta }^2}}{4
	    {f^2}}+\frac{{\delta f}}{f}\big)}^2}}}}\right\},
\label{eq:deltapz}
\end{equation}
	
\noindent{while} a transverse (lateral) offset by an amount
\(\delta\)x in the \(\xi\)-plane (Fig.~\ref{fig:feeddisplacement}, lower
half) will cause a path-length variation of 

\begin{equation}
{{{\delta p}}_x} = {\delta x} \frac{\xi
}{f}\left\{\frac{1}{1+\frac{{\delta f}}{f}} -
\frac{1}{{\sqrt{\frac{{{\xi }^2}+{{\eta
  }^2}}{{f^2}}+{{\left(1-\frac{{{\xi }^2}+{{\eta}^2}}{4
  {f^2}}+\frac{{\delta f}}{f}\right)}^2}}}}\right\}.
\label{eq:deltapx}
\end{equation}
			
In the reduction process of the holography data, these terms are found
by a fit of the measured beam map. The final map of surface deviations
is then referred to a position of the feed in the fitted
``out-of-focus'' location.

\section{Practical Application of the Holography Measurements}
\label{measurementdesc}

\subsection{Task}
\label{task}

\begin{figure}
  \resizebox{\hsize}{!}{
  \includegraphics[scale=1.00,angle=0]{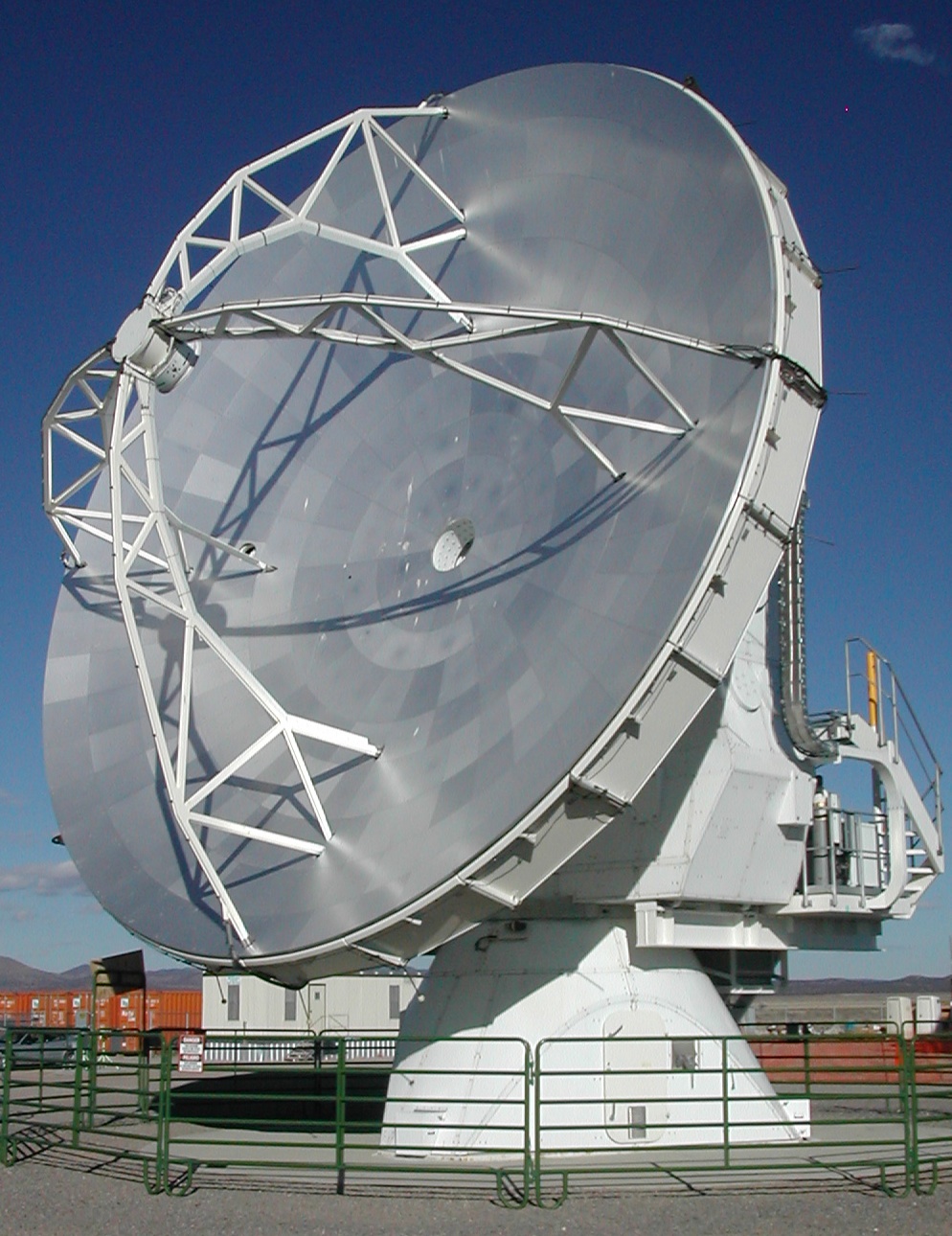}
  \includegraphics[scale=0.335,angle=0]{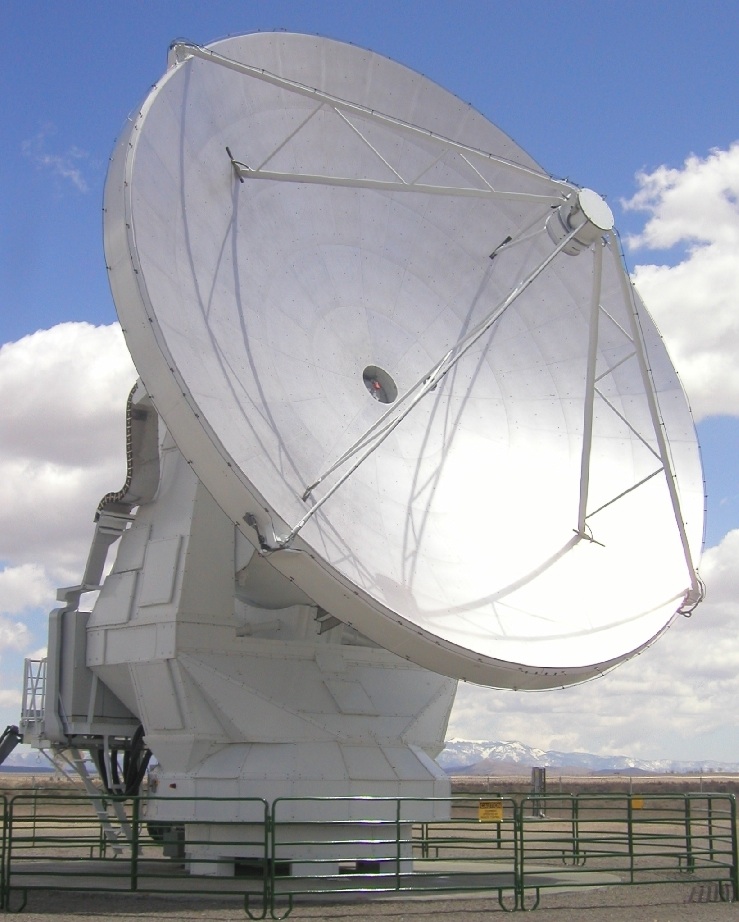}}
  \caption{The VertexRSI (left) and AEC (right) ALMA prototype antennas.}
  \label{fig:antennapics}
\end{figure}

We now describe the way in which a holography measurement has been
executed on the ALMA prototype antennas
(Fig.~\ref{fig:antennapics}). The specification requires the antennas
to have a surface 
accuracy of 25$\mu$m RMS for the AEC antenna (with a goal of 20$\mu$m)
and 20$\mu$m for the VertexRSI antenna. ALMA assumed the task to
demonstrate this with the aid of a holography system at 3~mm
wavelength after delivery of the antennas by the contractors with a
surface accuracy of not worse than 100$\mu$m RMS. This initial setting
was performed by VertexRSI with digital photogrammetry and by AEC with
the aid of a Leica ``total station'' laser-tracker (essentially a
theodolite with integrated distance measurement instrument and
all-electronic readout).

The holography system was designed to provide a measurement
repeatability of 10$\mu$m, which would suffice to demonstrate the realism
in the obtained overall surface accuracy. It should be noted that in
the current setup the holography system provides a surface map at one
elevation only. No information on the gravitational deformation of the
antenna with varying elevation angle can be obtained. 

\subsection{Equipment and Measurement Program}
\label{measprogram}

\begin{itemize}
\item The signal source for the holography measurements is a
  monochromatic transmitter at a frequency of 78.92 or 104.02 GHz,
  located on a 50 m high tower at a distance of 315 and 302 m from the
  VertexRSI and AEC antenna, respectively. The elevation angle is
  approximately 9 degrees. 
\item The receiver is a full-phase double-receiver, located in the
  apex region behind the primary focus of the main antenna. The
  reference signal is received by a wide beam horn pointing along the
  boresight towards the transmitter. 
\item Amplitude and phase maps of the antenna beam were obtained by
  raster scanning. The Nyquist sampling theorem provides the link
  between the angular size of the observed map and the required
  spatial resolution over the aperture. If we want to obtain n
  independent samples over the diameter of the aperture, we need to
  extend the map to an angle of n times the half-power beamwidth off
  axis. We chose a map size to obtain about 0.15 m spatial resolution
  after Fourier transformation of the map.  A typical measurement then
  takes about one hour of time.
\item From the phase distribution, which is a representation of the
  misalignment of the 264 (VertexRSI) or 120 (AEC) panels constituting
  the reflector, the necessary adjustments of the 5 support points per
  panel were derived. These were then applied by hand with a simple
  tool to improve the accuracy of the reflector surface. 
\end{itemize}

The algorithms and software used for the data analysis and derivation
of the panel adjustments have been applied successfully at the
telescopes of IRAM. The necessary corrections for the finite distance
to the transmitter (the ``near-field'' corrections) in our case were
derived and checked against similar corrections applied by others,
\eg\ for the JCMT \cite{Hills2002}.

The equipment has been designed to provide sufficient signal--to--noise
ratio to render the error due to noise insignificant. The greatest
risk in this type of measurement lies in undetected or poorly
corrected \textit{systematic errors}.

\begin{itemize}
\item An accurate knowledge of the amplitude and phase function of the
  feedhorn, illuminating the reflector, is essential, because errors
  in these are fully transferred to the aperture phase map and hence
  to the surface profile.
\item Multiple reflections from the ground or structures form a
  possible source of errors in this type of work. We carefully covered
  all areas of potentially harmful reflections with absorbing
  material. In some controlled experiments we could not demonstrate
  the existence of reflections.
\item The dynamic range of the receiver must be sufficient to
  accommodate the strong signal on the peak of the beam and the very
  weak signals towards the edge of the scan. There might have been
  some saturation on some of the measurements. We discuss this in more
  detail below.
\item The effect of the finite distance of the transmitter can be
  removed to a large extent (but not completely) by an axial shift in
  the position of the feed. An error in the distance to the
  transmitter thus can be corrected in the data analysis by a small
  adjustment of the feed position. The remaining phase error can be
  accurately calculated and applied to the data.
\end{itemize}

\subsection{Holography System Hardware}
\label{holohardware}

The hardware specifications and requirements are summarised in Tables
\ref{tab:holoreqs} and \ref{tab:holospecs}.  In the following we
briefly describe the hardware components that comprise the
holographic measurement system.

\begin{table}
\centering
\caption{Holography Hardware Requirements}
\label{tab:holoreqs}
\begin{tabular}{|l|l|}
\hline
Measurement Error & $<10\mu$m \\
Phase Accuracy & $<0.3\deg$ ($2.5 \mu m$ @ 3mm) RMS \\
Amplitude Accuracy & $<1\%$ \\
Dynamic Range & $\geq43$dB \\
Signal-to-Noise Ratio (SNR) & $\geq40$dB \\
Channel-to-Channel Isolation & $>100$dB \\
Data Rate & $\sim 80$ samples/sec (12 msec)\\
\hline
\end{tabular}
\end{table}

\begin{table*}
\centering
\caption{Holography Hardware Specifications}
\label{tab:holospecs}
\begin{tabular}{|l|l|}
\hline
Frequencies & 78.92 and 104.02 GHz \\
Frequency Stability & $\leq\pm5$ Hz/day \\
Receiver Bandwidth & 10 kHz \\
Receiver Tunability & 130 MHz \\
Transmitter Antenna Gain & 33dB \\
Transmitter EIRP (P) & $>20\mu$W \\
Transmitter Power to Antenna & $>10$nW \\
Transmitter Antenna Beam-Width @ $-3$dB & $4.6\deg$ (twice antenna angle
at xmtr) \\
Reference Antenna Beam-Width @ $-3$dB & $4.6\deg$ (twice scan range) \\
Main Feed Beam-Width @ $-3$dB & $128\deg$ ($-3$dB edge taper) \\
System Temperature & 3200 K \\
Reference Feed Power Received ($P_r$) & $1.7\times10^{-9}P$ \\
On-Boresight Signal ($M_0$) & $4.2\times10^{-7}P$ \\
On-Boresight Noise ($\sigma_0$) & $\left(1.2\times10^{-22}W (P)\right)^\frac{1}{2}$ \\
Off-Boresight Noise ($P_r$ Term) & $\left(2.1\times10^{-27}W (P)\right)^\frac{1}{2}$ \\
Average map noise for complex correlator ($\sigma_{av}$) & $\left(2.2\times10^{-25}W (P)\right)^\frac{1}{2}$ \\
\hline
\end{tabular}
\end{table*}

\subsubsection{Front-End}
\label{holofe}

\begin{figure}
  \resizebox{\hsize}{!}{
  \includegraphics[scale=0.20]{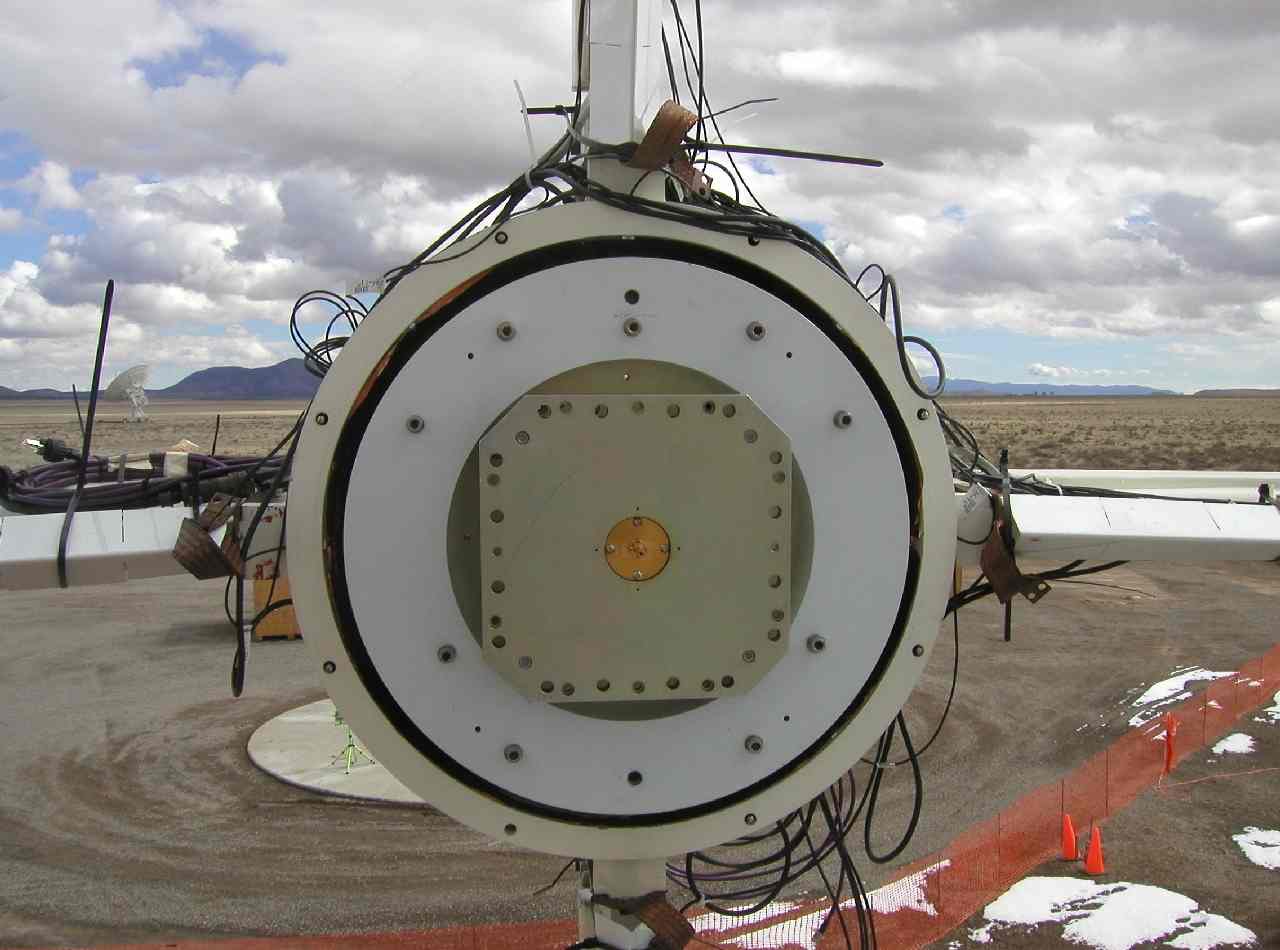}\\[10pt]
  \includegraphics[scale=0.20]{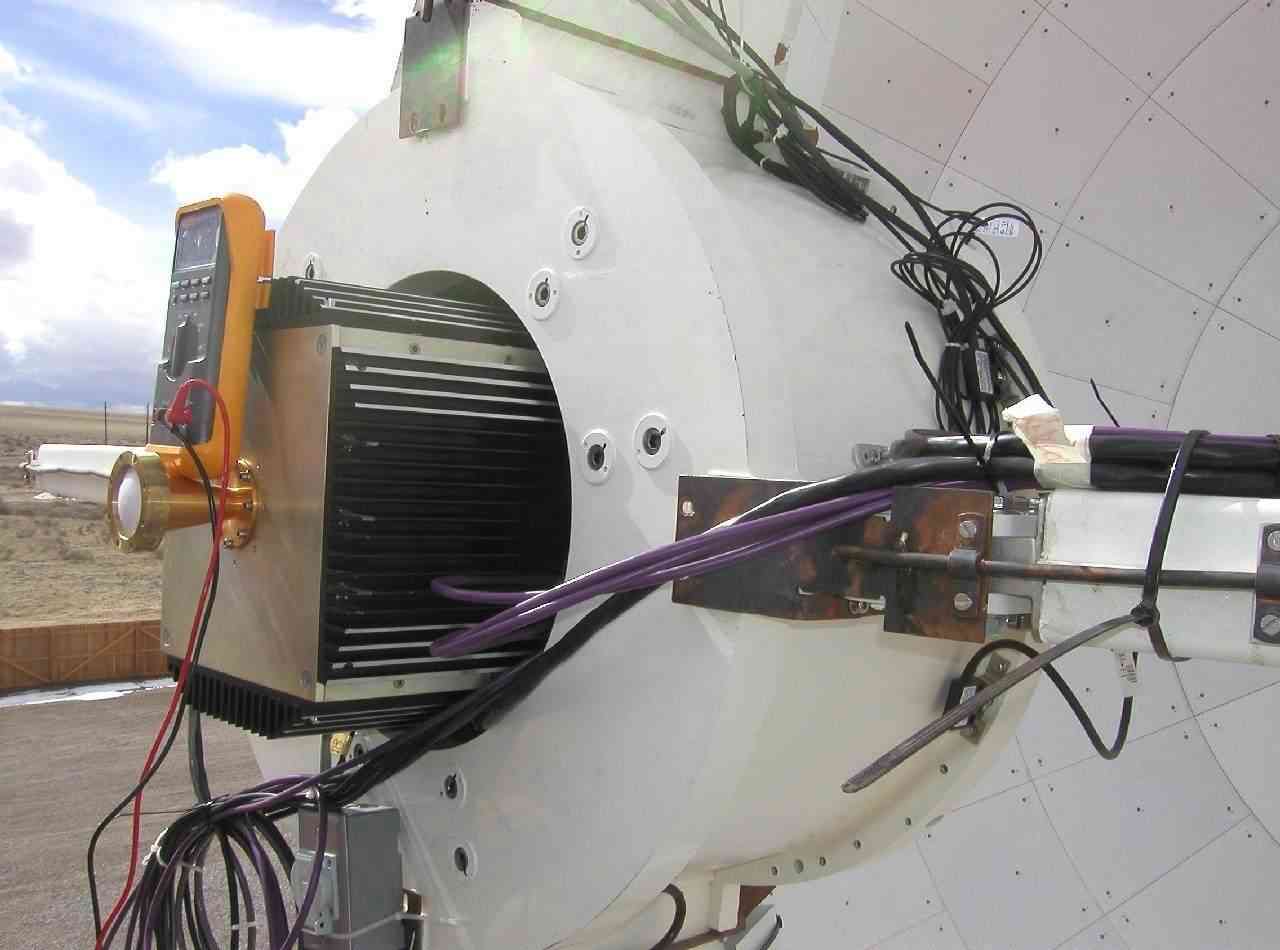}\\[10pt]
  \includegraphics[scale=0.20]{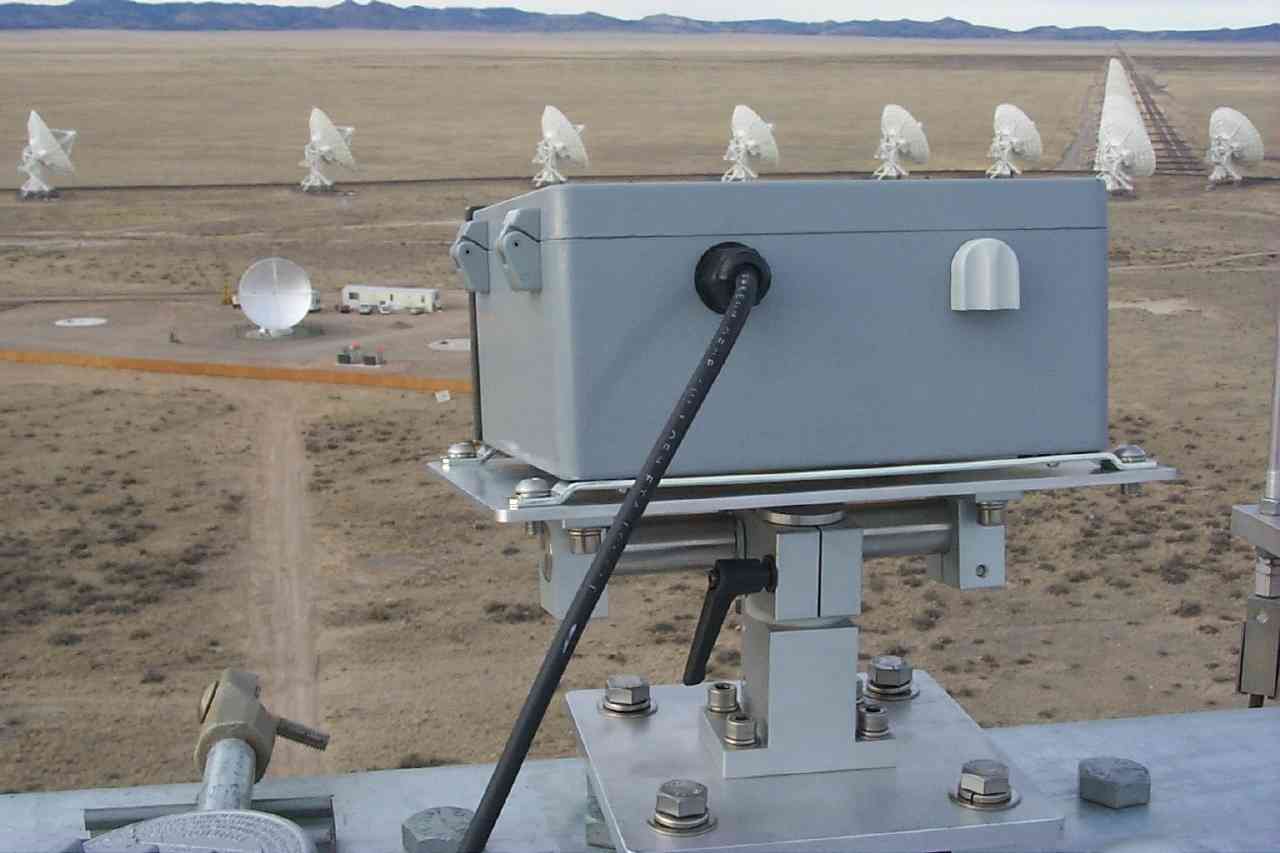}}
  \caption{Holography system hardware.  Left: Signal feed side of the
  front-end. Middle: Reference feed side of the front-end.  Right:
  transmitter on top of the tower pointing at the VertexRSI prototype 
  antenna in the foreground.  The other antennas are part of the NRAO
  Very Large Array (VLA).}
  \label{fig:holofe}
\end{figure}

The front-end (see Fig.~\ref{fig:holofe}) is enclosed
in a small, temperature controlled box with a diameter of about 30 cm
and a length of 50 cm. It fits inside the ``apex structure'' behind the
primary focus of the VertexRSI antenna. The AEC antenna does not
provide such a wide space and the receiver is bolted to the outside
flange of the apex structure with a long piece of waveguide bringing
the signal feed in focus. Both the signal-- and reference--receiver are
housed ``back-to-back'' in this box. This provides a compact system in
which the LO signals can easily be made equal in length, greatly
contributing to the phase stability of the system. Broadband mixers at
ambient temperature convert the received signal frequency to a
baseband of 10 kHz width. The system is designed with two frequencies
at 78.9 and 104.02 GHz. Making the measurement at two different
frequencies can be helpful in discerning systematic effects in the
resulting maps, for instance caused by multiple reflections. The
receiver is also tunable around each of these frequencies by 130 MHz
for similar reasons. The signal horn is a conical, grooved
cylindrical waveguide horn, while the reference horn is of similar
design and equipped with a lens to provide a reference beam with a
beam-width of 4.6 degrees at the half-power points. 

As is clear from the theoretical treatment above, it is imperative
that we know the amplitude and phase function of both the reference
and the signal feed as accurately as possible. The phase function must
be subtracted from the measured aperture phase before connecting its
phase variations to errors in the reflector profile. The feedhorns
have been measured with great care on the indoor range at IRAM in
Grenoble \cite{Lazareff2003}. The results were compared with model
calculations using an advanced electro-magnetic simulation package
(the FDTD package of Microwave Studio from Computer Simulation
Technology) and excellent agreement was found. The phase pattern of
the feeds have an
estimated error of less than one degree,  while the amplitude taper at
the edge of the reflector aperture is $-6$~dB. This is more than we
would like (a free--space taper of 2.5 dB has to be added to the
measured level).  For a high signal to noise ratio in the outer part of
the reflector an actual level of $-$6~dB is preferred. For the
measurement of the ALMA production antennas this feed should be
replaced by one which provides such a taper. 

\subsubsection{Back-End and Transmitter}
\label{holobetx}

The back-end of the receiver is essentially a digital signal processor
(DSP) where the narrow-band signals are digitized and correlated. Both
the ``sine'' and ``cosine'' part of the complex correlation function are
obtained, which are then transformed to the amplitude and phase
functions.

The transmitter consists of a single photo-diode, directly coupled to
a waveguide horn, which is fed through an optical fiber by two optical
signals at different frequencies near a wavelength of $\sim 1550$
nm. The photo-diode provides a mixing signal at the difference of the
two optical signals, tunable roughly from 78.7 to 79.0 GHz (low band)
and 103.8 to 104.2 GHz (high band), with an
output power of about 10 nW, leading to an EIRP of about 20
$\mu$W. The transmitter is placed on top of a 50 m high tower at a
distance of 300 to 325 m from the aperture of three antennas at the
site, resulting in a measurement elevation angle of about 9 degrees.

\section{Holographic Data Acquisition}
\label{acquisition}

To derive typical values for the various holography map parameters,
we set the following boundary conditions:

\begin{itemize}
\item The data rate is the canonical 12 msec per sample, which
means about 80 samples per second.
\item The fine tuning feature of the holography receiver allows for
the search for ground reflection.
\item A goal for the total time for one map is less than one hour.
\item The required aperature plane resolution is $\le$ 20 cm.  This
yields $\ge$ 25 independent points per square meter of reflector
surface. 
\item The 5 panel support points are on average some 0.4 to 0.6 m
  apart for the VertexRSI and AEC panels, respectively. A twist in the
  panel with a similar scale length can be partially corrected. With a
  measurement spatial resolution of 15--20 cm this large scale twist
  can be fitted sufficiently well.
\item Oversample by a factor of at least 2 to minimize aliasing.
\end{itemize}

Based on the equations listed in Appendix \ref{equationsandcalcs},
with ${f_1} = 1.13 (6 + 2.5~\textrm{dB taper})$, we obtain the typical
holography map parameters of Table \ref{tab:holoparms}.

\begin{table*}
\centering
\caption{Typical Holography Map Parameters}
\label{tab:holoparms}
\begin{tabular}{|l|l|l|l|l|l|l|l|l|}
\hline
Map Type & $\delta_d$ & $f_{osr}$ & $\theta_{ext}$ &
$\theta_{sr}$ & $\dot{\theta}$ & 
$N_{row}$ & $f_{oss}$ & $t_{map}$ \\
& (cm) && (deg) & (arcsec) & (arcsec/s) &&& (hr) \\ \hline

Standard & 20 & 2.2 & 1.64/1.24 & 33/25 & 300 & 180 & 20/15 &
0.96/0.73 \\

Fine & 13 & 2.2 & 2.46/1.87 & 33/25 & 600 & 270 & 40/30 &
1.08/0.82 \\

Course & 20 & 1.4 & 1.64/1.24 & 53/40 & 300 & 112 & 20/15 &
0.61/0.46 \\

\hline
\multicolumn{9}{l}{Assumes $f_1$ = 1.13 (6 + 2.5 dB taper), $\nu$ =
  78.92/104.02 GHz, $\theta_b$ = 74/56 arcsec, and $f_{apo}=1.3$.} \\
\end{tabular}
\end{table*}

\section{Holographic Data Analysis}
\label{analysis}

\subsection{Description}
\begin{figure}
  \resizebox{\hsize}{!}{
  \includegraphics{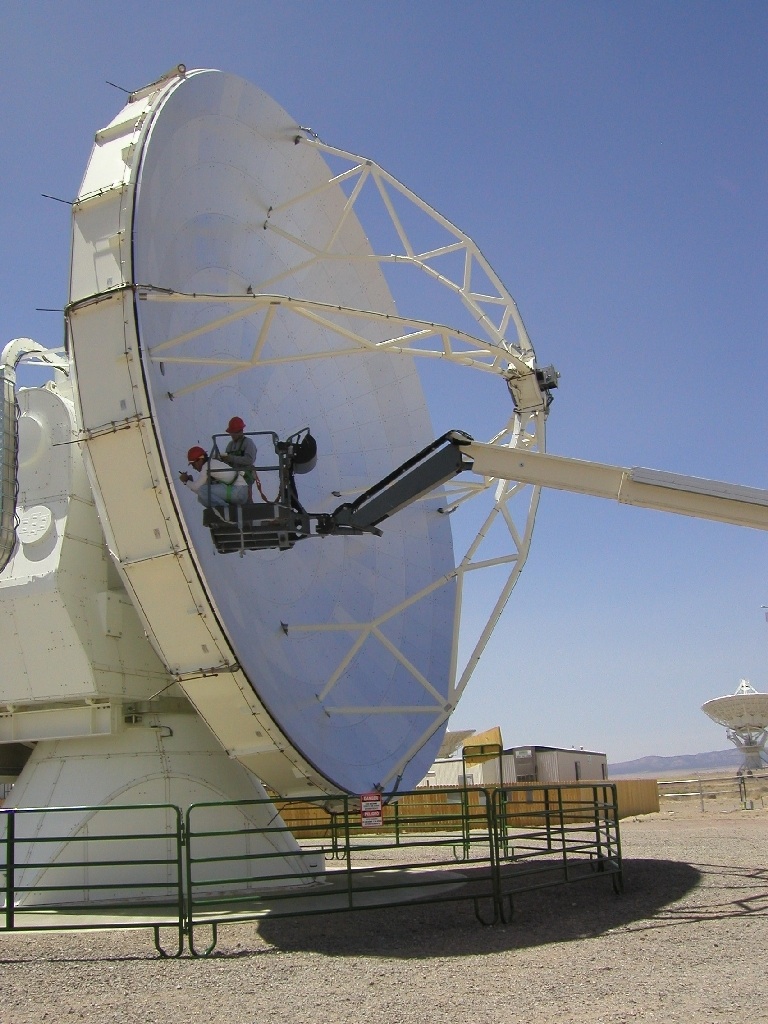}\\
  \includegraphics{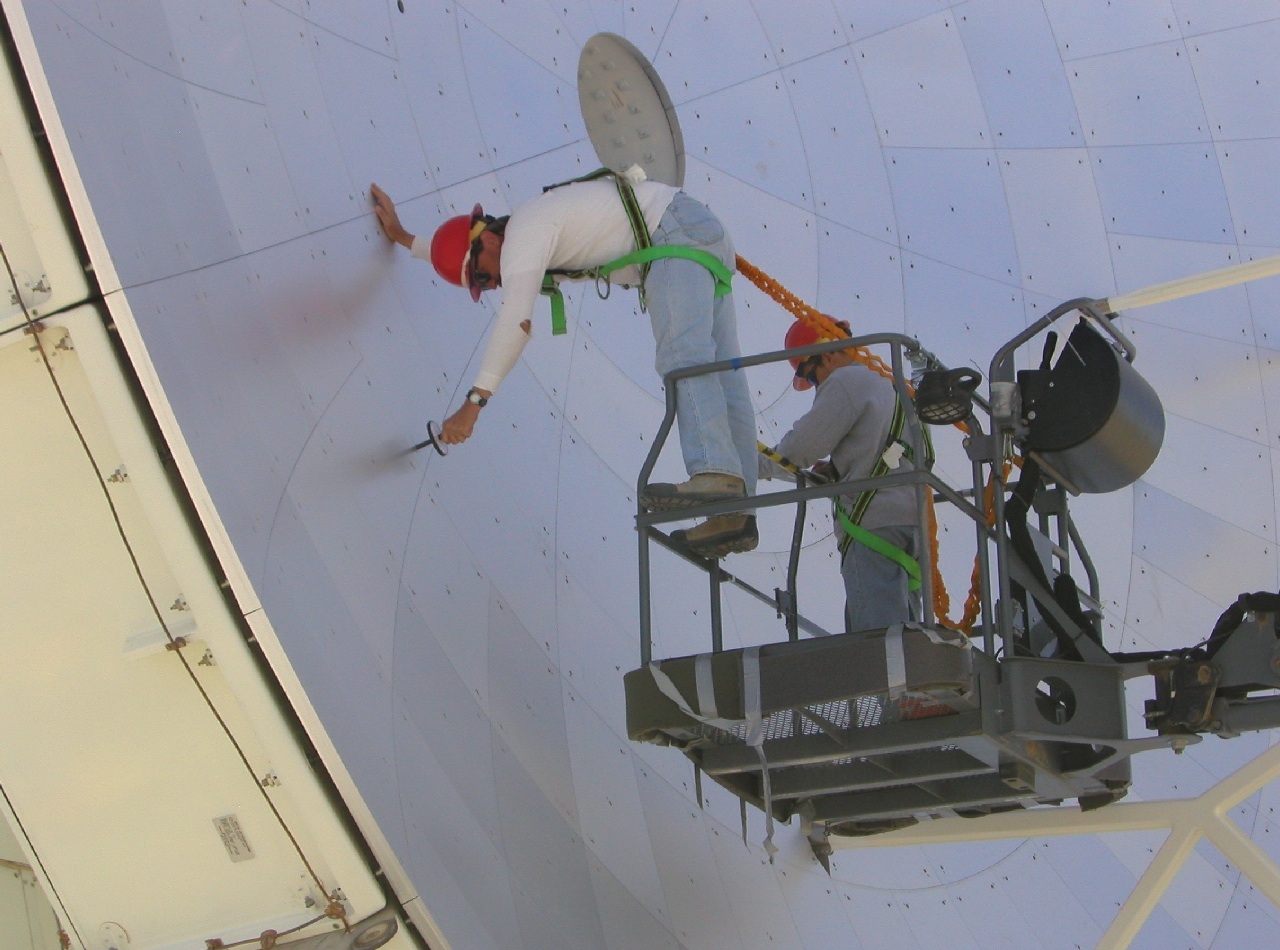}}
  \caption{Panel adjustment of the VertexRSI prototype antenna.}
  \label{fig:holosetting}
\end{figure}

Data analysis uses the CLIC data reduction software of the Plateau de
Bure interferometer.  The raw data, written by the on-line software
in the ALMATI-FITS data format \cite{Lucas2001}, is converted to
Plateau de Bure format using CLIC. The data are then calibrated and
imaged using CLIC. The two main operations are: 

\begin{enumerate}

\item {\em Calibrate data in amplitude and phase}, based on boresight
 measurements at beginning and end of each map row, assuming gradual
 drift in amplitude and phase with time. This uses the standard
 amplitude and phase calibration commands in CLIC, which:

 \begin{enumerate}
 \item Fit cubic spline functions of time to the observed amplitude
   and phase data on the boresight measurements.
 \item Subtract the phase spline function from the observed phase for
   the mapping scans.
 \item Divide the observed amplitude for the mapping scans by the
   amplitude spline function.
 \end{enumerate}

\item {\em Compute the aperture map and fit panel displacements and
 deformations}.  The data processing steps for computation of the
 aperture maps are:

 \begin{enumerate}

  \item {\em Interpolate data to a regular grid} in the antenna-based
  coordinate system. This grid matches the observed system of rows
  (same number and separation). This grid is further extended, by
  addition of zeroes, to a user-specified size, in order to get a
  finer interpolation of the output aperture map: 64x64, 128x128,
  256x256 and 512x512 sizes are available.

  \item {\em FFT to aperture plane}. This is replaced by a more complex
 transformation if one takes into account the first non-Fresnel
 terms, as described in \S~\ref{sec:fresnel}
 (Eq.~\ref{eq:fchieta3}).

 \item {\em Compute phases in the aperture plane}.

\item {\em Apply the geometrical phase correction}:
 \label{sec:correction} This is Eq.~\ref{eq:deltap1} plus
 Eq.~\ref{eq:deltap2}, substituting $\rho = \sqrt{\xi2+\eta2}$
  as the radius in the aperture, $f$ the focal length of the primary,
  and $\delta f$ by the distance between the
  holographic horn phase center and the antenna prime focus (see
  \S\ref{sec:fresnel}).

 \item {\em Correct for measured feed phase diagram}. 

  The measurement is described in the memo by Lazareff \etal
  \cite{Lazareff2003}.

 \item {\em Mask edges and blockage}.

 \item {\em Fit and remove 6 phase terms:} Constant, 2 linear
   gradients in the horizontal and vertical directions, 3 focus
   translations. These terms account for a phase offset, an antenna pointing
   error (constant during the measurement) and a small vector
   displacement of the holography horn relative to the nominal focus
   position ($f+\delta f$). The phase terms for the axial and
   transverse displacements of the focus are given in
   Eqs~\ref{eq:deltapz} and \ref{eq:deltapx} respectively. Optionally
   one may keep fixed either the $\delta x$ and $\delta y$ coordinates
   or all three $\delta x$, $\delta y$, and $\delta z$
   coordinates. 

 \end{enumerate}

 \item {\em Convert to normal displacement map.}  See Appendix \ref{rmscalc}
 for details.

 \item {\em Plot amplitude and phase maps.}

 \item {\em Fit panel displacements (optionally deformations) and
   screw adjustments}. 

   \begin{enumerate}

   \item Each panel is assumed to be displaced (in the
     axial direction), tilted (around two orthogonal axes), and
     possibly deformed (deformation is a quadratic function of
     position offset relative to the panel center). As there are only
     five screws, only two deformation modes are allowed, we have thus
     five displacement modes:
     \begin{eqnarray}
     \delta p_1(x,y) & = & a \\
     \delta p_2(x,y) & = & b x \\
     \delta p_3(x,y) & = & c y \\
     \delta p_4(x,y) & = & d (x^2 + y^2) \\
     \delta p_5(x,y) & = & e (x^2 - y^2) 
     \end{eqnarray}
       where $x$ and $y$ are coordinates of a point on a panel ($x$
       and $y$ in the plane tangent to the panel surface, with $x$
       axis in the radial direction). 

     \item The five coefficients $a,b,c,d,e$ are independently fitted
       for each panel to the relevant part of the map for this panel.

     \item To take into account the effects of finite angular
       resolution, an iterative procedure is used where:

       \begin{enumerate}
       \item The radiated beam is calculated using the fitted panel
       surfaces.
       \item This beam is truncated to the size of the observed beam
	 map, and subtracted from the observed beam.
       \item An incremental surface map is calculated.
       \item A new set of incremental panel displacements is
	 calculated from this map.
       \end{enumerate}

       The procedure converges after a few iterations.  The screw
   settings are output to a text file.

  \end{enumerate}
 \item {\em These screw settings are applied to the panel adjusters}
   to improve the surface accuracy of the reflector.  The adjustments
   were done with a simple tool.  Two people on a manlift approached
   the surface from the front, where the adjustment screws are
   located (see Fig.~\ref{fig:holosetting}).  The time needed for an
   adjustment of the total of 1320/600 adjusters was 8/7 hours for the
   VertexRSI/AEC prototype antennas, respectively.  The entire
   procedure is repeated until the required accuracy is achieved, or
   alternatively until the inherent measuement accuracy has been
   reached and no further improvement is surface accuracy is obtained.
   
\end{enumerate}

\section{ALMA Prototype Antenna Holography Measurement Results}
\label{results}

\begin{figure}
\resizebox{\hsize}{!}{
\includegraphics[angle=-90]{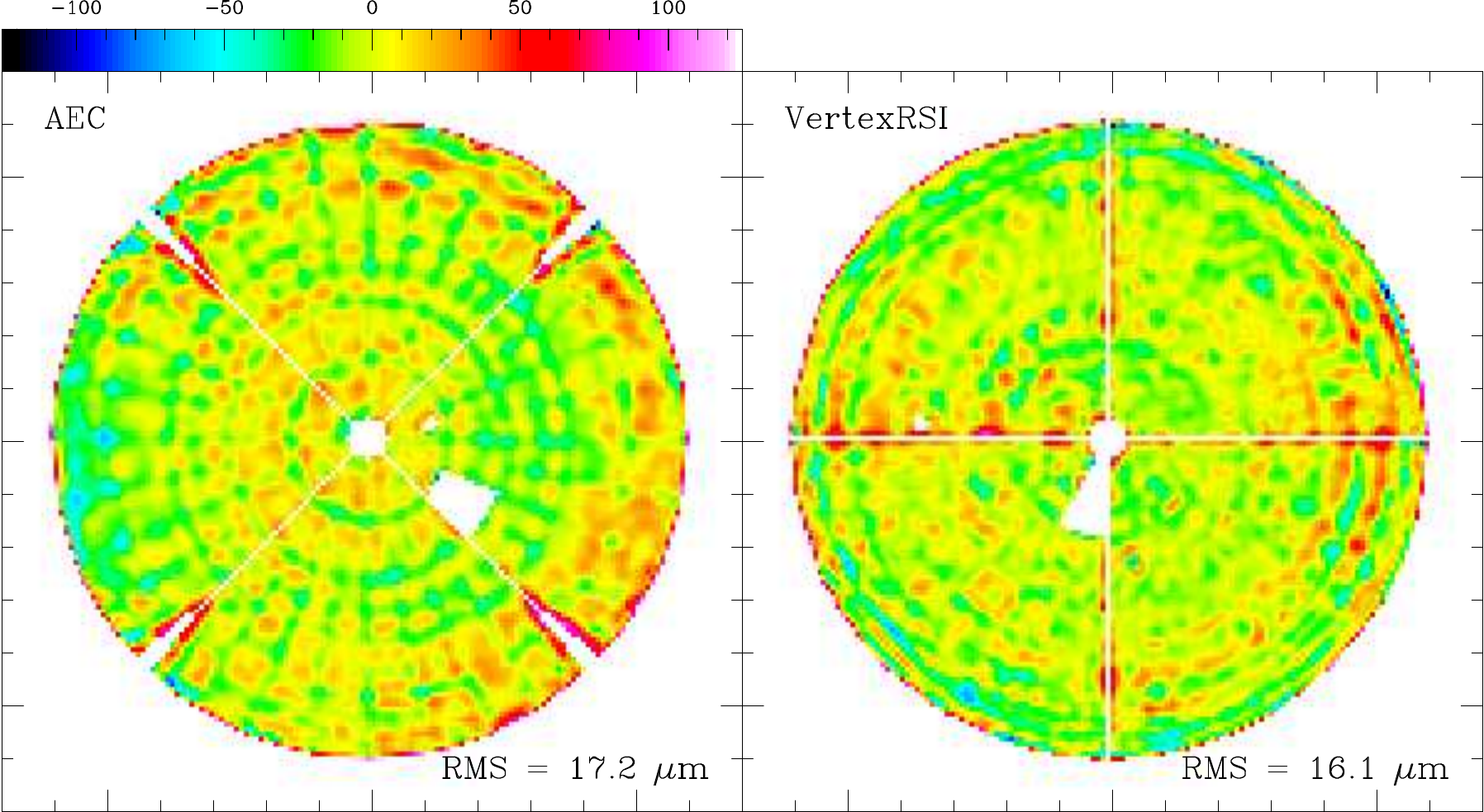}}
\caption{Final holography maps for the AEC (left) and VertexRSI
  (right) ALMA prototype antennas.  Horizontal and vertical axes units
  are meters with color scale plotted as surface displacement in
  $\mu$m (top right).  Note that the ``$\times$'' 
    (AEC) and ``+'' (VertexRSI) feed leg structure and the
    difficulties encountered with holographic measurements near these
    structures leads to the poor measurement results in these areas.}
\label{fig:holocompfinal}
\end{figure}

The holographic measurement and setting of both antennas was performed
immediately after the antennas became available for evaluation. During
a period of about one year the antennas were subjected to a number of
hard loads, like fast switching tests, drive system errors resulting
in strong vibrations, and high-speed emergency stops collisions. Also,
the influence of wind and diurnal temperature variations on the
surface stability was a point of concern. It was therefore decided to
close the evaluation program with a second holographic measurement of
the reflector surfaces. This was done in December 2004 to February
2005 during relatively good atmospheric conditions. It was during
these measurements that we discovered that we had not correctly taken
care of the correction explained above in \S\ref{aperref}. This
correction was subsequently applied properly and the final surface
maps are shown in Fig.~\ref{fig:holocompfinal}. Both antennas
were set to a surface accuracy of 16--17 $\mu$m RMS.

\subsection{VertexRSI Antenna}
\label{vrsiresults}

\subsubsection{Overview}
\label{vrsioverview}

The antenna was delivered with a nominal surface error of 80 $\mu$m
RMS, as determined from a photogrammetric measurement. Our first
holography map showed an RMS of approximately 85 $\mu$m. A first
setting of the surface resulted in an RMS of 64 $\mu$m. In four more
steps of holographic measurement followed by adjustment the surface
error decreased to 19 $\mu$m RMS.

\begin{figure}
\resizebox{\hsize}{!}{
\includegraphics[scale=0.75]{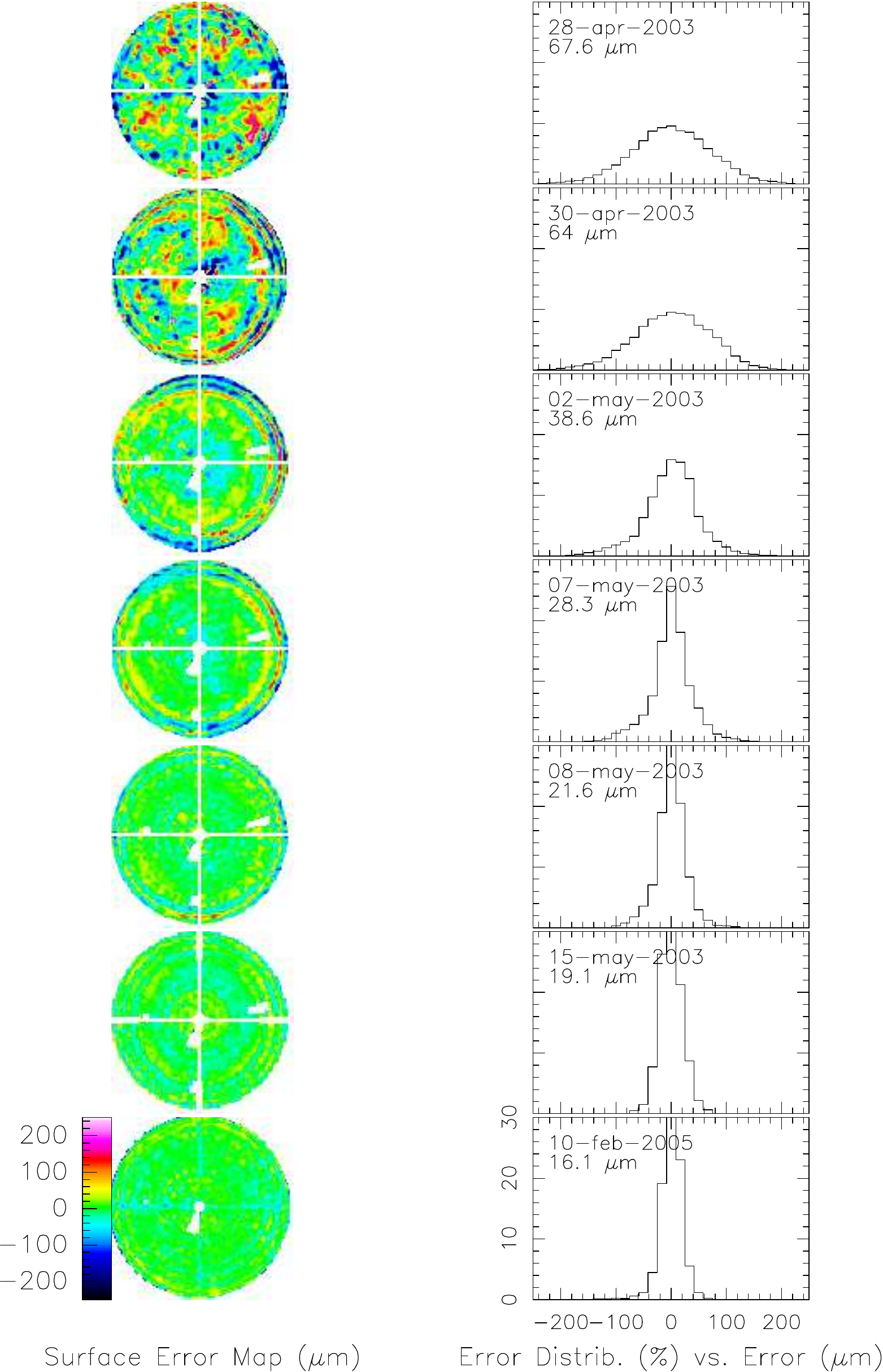}}
\caption{Sequence of surface error maps with intermediate panel
  setting. The surface contours are shown on the left side; the error
  distribution, plotted as percentage contribution versus surface
  error, on the right. The white cross and the small white areas
  represent the quadripod and a few faulty panels and were not
  considered in the calculation of the RMS error.  Progression to the 
  final surface RMS included holography system checkout, so does not
  represent the time required to set the VertexRSI antenna to its
  final surface.}
\label{fig:holosVertex}
\end{figure}

The sequence of surface error maps, along with the RMS and the error
distribution is shown in Fig.~\ref{fig:holosVertex}.
As allowed in the specification, we have applied a weighting over 
the aperture proportional to the illumination pattern of the feed. 
This essentially diminishes the influence of the surface errors in 
the outer areas of the reflector. The white areas in the surface error
maps are the quadripod, optical pointing telescope, and a few bad
panels, which could not be set accurately.  All of these structures
were left out of the calculation of the final overall RMS value.

With increasing accuracy the presence of an artefact in the outer area
of the aperture became apparent. There is a ``wavy'' structure in the
outer section with a ``period'' too large to be inherent in the
panels. Experiments with absorbing
material showed that it was not caused by multiple reflections. The
effect can be described by a DC-offset in the central point of the
measured antenna map, \ie\ some saturation on the point with the
highest intensity. By adjusting this offset in the software, most of
the artefact could be removed.  This has been done with the final
data.  The additional set of follow-up holography measurements in
December 2004 -- February 2005 did not suffer from this signal
saturation, and no artefact was observed in these follow-up maps.
Checks of the holography hardware indeed suggest that the 2003
holography measurements did experience a small amount of signal
saturation.

The best surface maps were obtained at night. During the spring 2003
period they consistently show an RMS of about 20$\mu$m. Daytime maps
tend to be somewhat worse; typical values of the RMS lie between 20
and 25$\mu$m. Part of this is certainly due to the atmosphere, even
over the short path-length of 315 m.

\subsubsection{General Surface Stability}
\label{vrsisurfstab}

\begin{figure}
\resizebox{\hsize}{!}{
\includegraphics[scale=0.65,angle=-90]{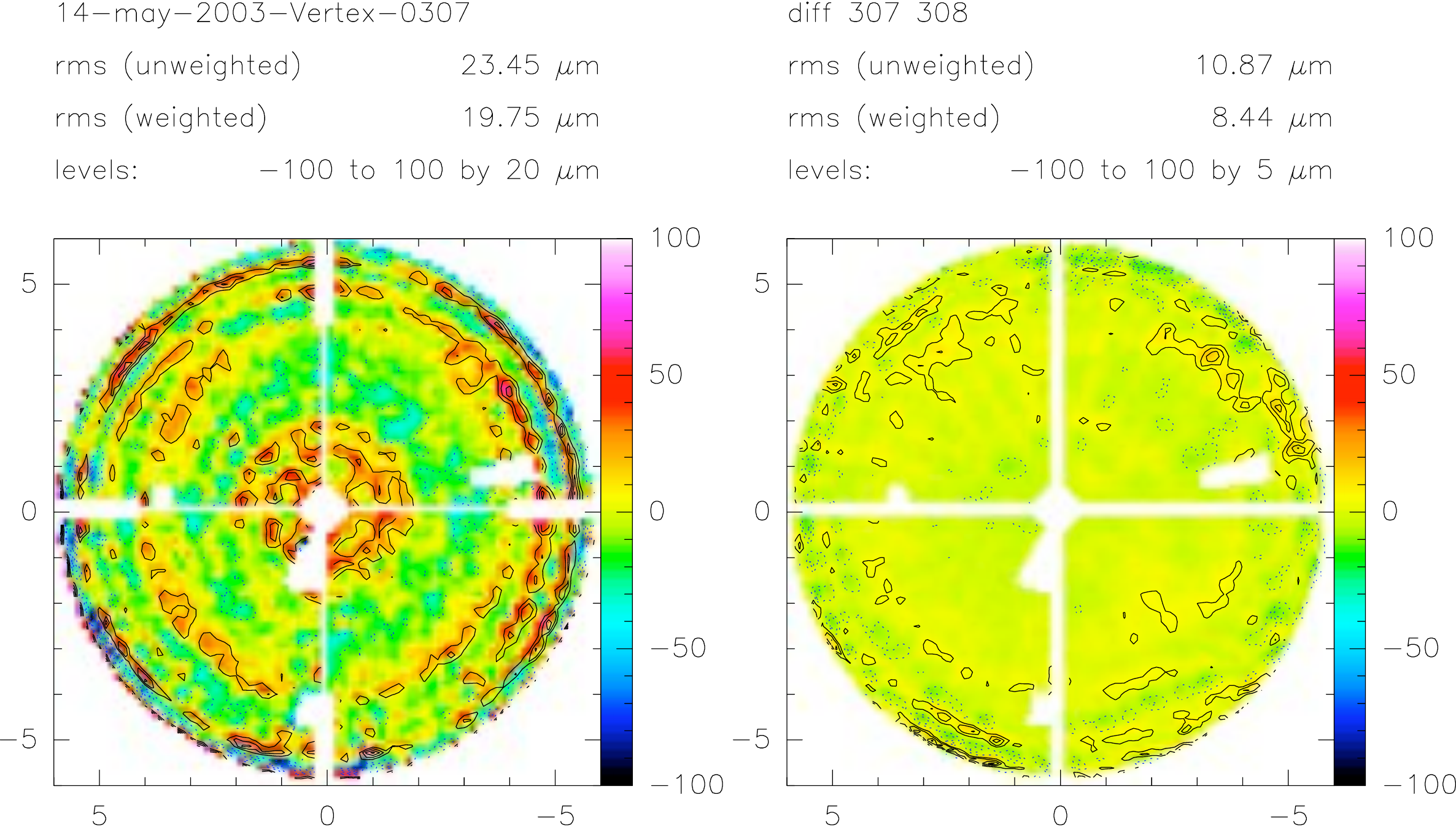}}
\caption{Example of the repeatability of the measurements. Horizontal
  and vertical axes units are meters with color scaling showing
  surface error in $\mu$m.  The map on the right is the difference
  between the one at left and a map made one hour afterwards. The RMS
  of the difference maps is about 8$\mu$m, which is commensurate with
  the expected value due to noise and atmospheric fluctuations.}
\label{fig:VertexDiff}
\end{figure}

To estimate the accuracy and repeatability
of the measurements, we produced difference maps between successive
measurements throughout the measurement period.  The RMS difference
between consecutive maps 
is normally less than 10$\mu$m, typically 8$\mu$m. An example of a
difference map is shown in Fig.~\ref{fig:VertexDiff}.  The map of
measurement number 307 is shown on the left, while the right hand side
shows the difference between map 307 and 308, made about one hour
later.

We have made many maps after the final setting of the surface while
changing the orientation of the antenna with respect to the Sun. 
Maps taken over a few consecutive days were obtained at widely
different temperatures, while also the wind conditions varied over
time. The measured rms error during a 30 hour period in early May 2003
varied between 20 and 23$\mu$m. During this period the wind was calm
($<5$ m/s) and temperature changes of 15 C were encountered. A later 5
day series in mid June gave rms errors from 22-26$\mu$m with
temperature variation up to 20 C, wind velocities up to 10 m/s and
periods of full sunshine. Most of this increase is believed to be due
to the deteriorating atmospheric conditions at the VLA site during
summer, when the humidity is significantly higher than normal. The
much better results of 17$\mu$m obtained during the cold and dry
winter period also point to a significant atmospheric component in the
spring and summer results. 

However, some of the changes will be caused by temperature and
wind. To increase the rms from 20 to 22$\mu$m, the ``additional''
component has a magnitude of 9$\mu$m rms. Such a contribution can be
expected from the calculated values of 4$\mu$m each for wind and
temperature for the panels, and 5$\mu$m for wind and 7$\mu$m for
temperature for the BUS. These numbers are all within the
specification. Actually, the measured differences are close to those
expected from the estimated accuracy of the holography measurement and
the measured rms differences in consecutive maps of about 8$\mu$m.

\subsection{AEC Antenna}
\label{aecresults}

\subsubsection{Overview}
\label{aecoverview}

The apex structure of the AEC antenna does not enable us to mount the
holography receiver inside the cylinder, as in the case of the VertexRSI
antenna. Thus in this case the receiver was bolted to the flange on
the ``outside'' of the apex-structure. Consequently, the feedhorn was
brought to the required position by a piece of waveguide of about
500~mm length. This caused significant attenuation in the received
signal from the reflector to the mixer. Considering the available
transmitter power, we concluded that this would not jeopardise our
measurement accuracy significantly.

The AEC antenna surface was set by the contractor with the aid of a
Leica laser-tracker. The RMS of the surface was reported by the
contractor to be 38$\mu$m. After this measurement a servo error caused
the elevation structure to run onto the hard stops at high speed. 
The contractor decided to repeat the surface measurement and 
obtained an RMS of 60$\mu$m with some visible ``astigmatism'' in the
surface.

\begin{figure}[h!]
\resizebox{\hsize}{!}{
\includegraphics[scale=0.80]{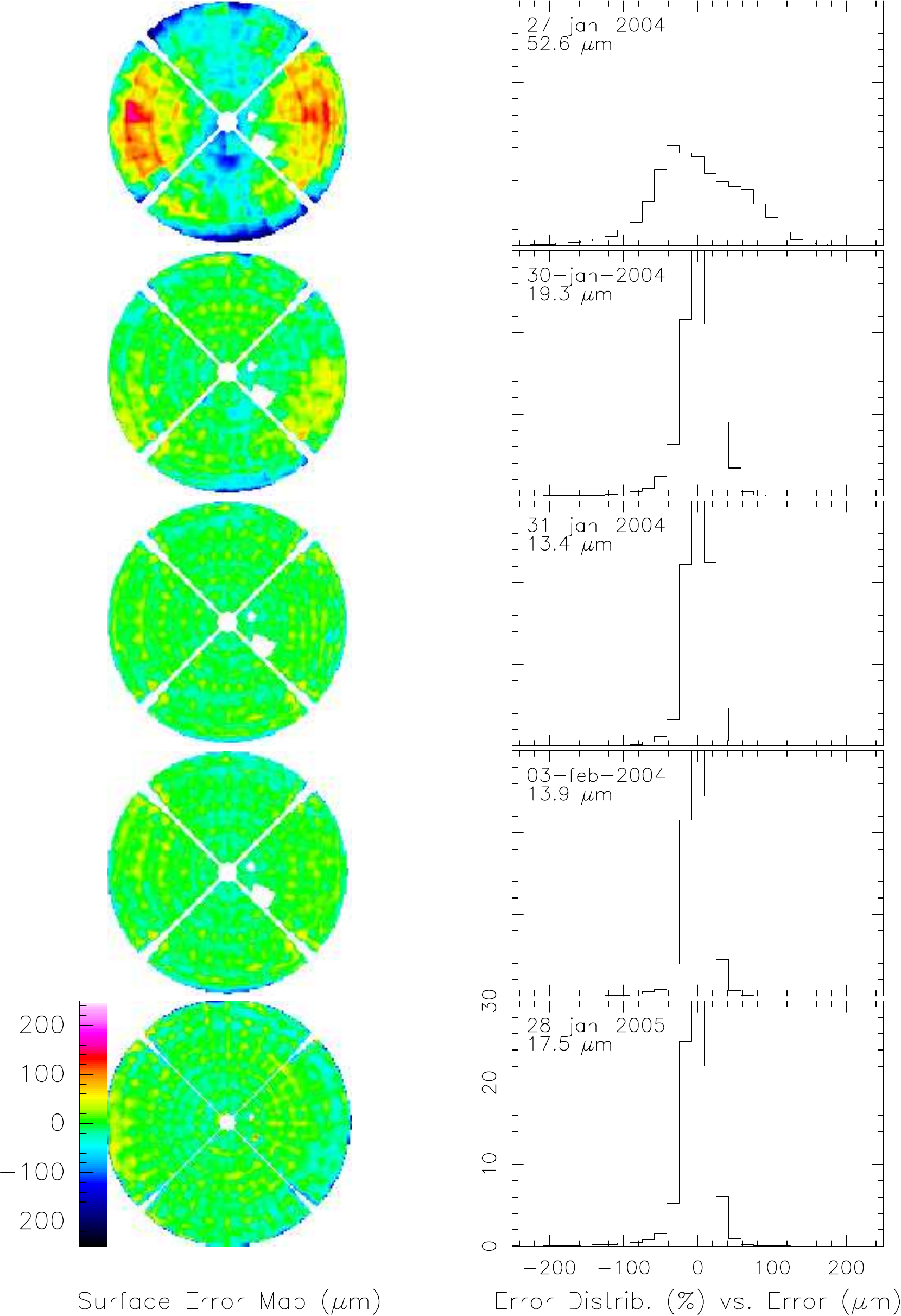}}
\caption{Sequence of surface error maps with intermediate panel
  setting. The surface contours are shown on the left side; the error
  distribution on the right. The white cross and the small white area
  represent the quadripod and a faulty panel and were not considered
  in the calculation of the RMS error.  Note that the differences
  between the final three maps is within the overall uncertainty of
  the holography measurements.} 
  \label{fig:holosAEC}
\end{figure}

Our first holography map indicated an RMS of 55$\mu$m with a clearly
visible astigmatism. We could identify the high and low regions with
those on the final AEC measurement. With two complete
adjustments we surpassed the goal of 20$\mu$m. A third partial adjustment
improved the surface RMS to about 14$\mu$m. There is no indication of
the ``artefact'' seen in the VertexRSI antenna. The results of the
consecutive adjustments are summarised in 
Fig.~\ref{fig:holosAEC}.  The last panel in this figure shows the
final map after the repeated measurement and setting in January 2005.

The adjustments were done with a tool provided by the contractor. It
was similar to the one used by us on the VertexRSI antenna, but it was
calibrated in ``turns'' rather than in micrometres. 

\subsubsection{General Surface Stability}
\label{aecsurfstab}

\begin{figure}
\resizebox{\hsize}{!}{
\includegraphics[scale=0.65,angle=-90]{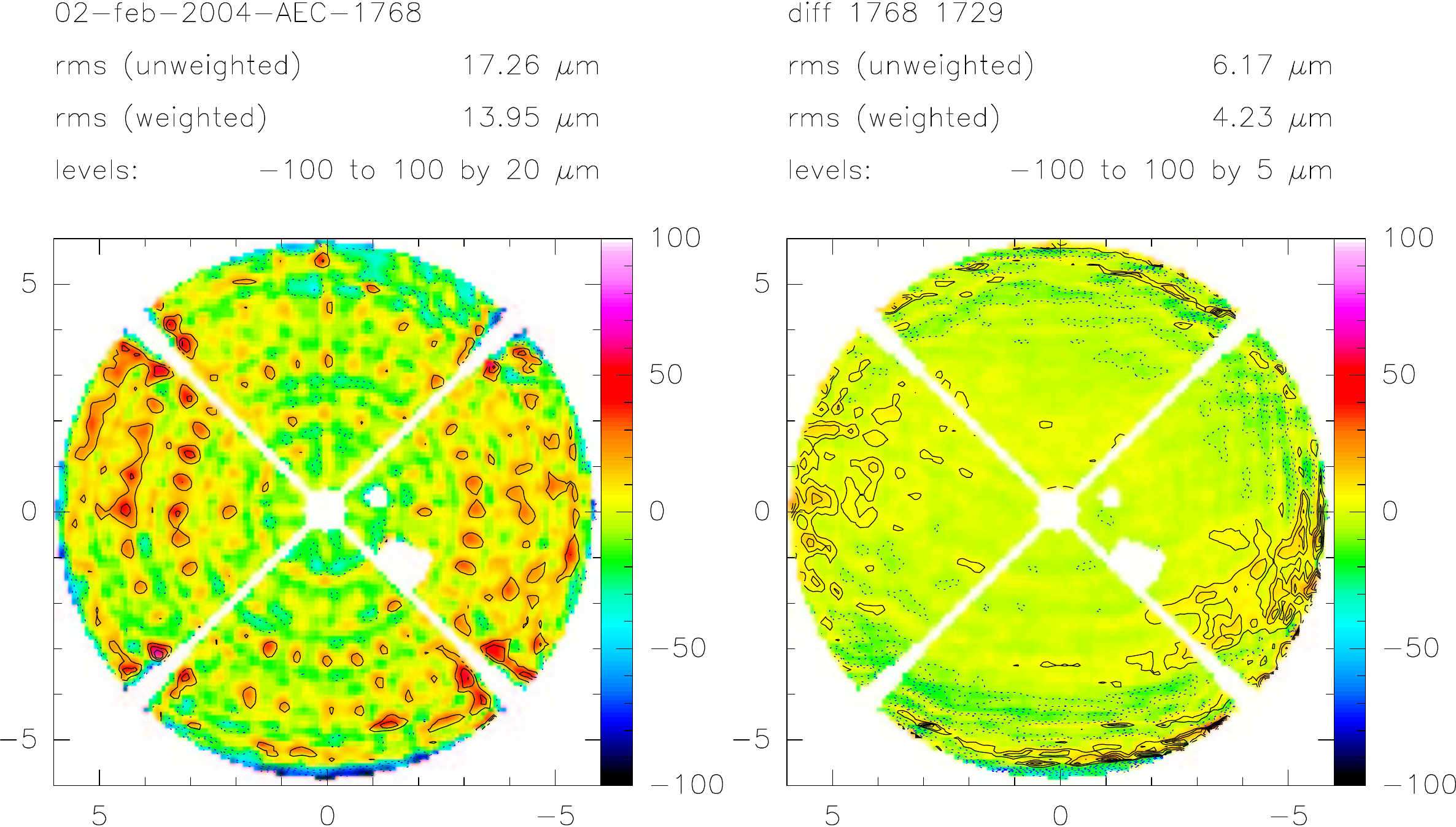}}
\caption{Example of the repeatability of the measurements. The map on
  the right is the difference between the one at left and a map made
  one hour afterwards. The RMS of the difference maps is about $5\mu$m.}
\label{fig:AECDiff}
\end{figure}

In Fig.~\ref{fig:AECDiff} we show one of the final results and a
difference map of this and the following measurement, made one hour
later. The difference map indicates a repeatability of $\sim 5 \mu$m
RMS. There is no indication of the ``ringing'' in the outer region of
the aperture, as was the case for the VertexRSI antenna. We ascribe
this to the lower signal level due to the long piece of waveguide
between feed and mixer.

Also here we made a series of 16 maps over a period of more than two
days in early February 2004. Temperatures ranged from $+2$ to $-10$ C,
while the wind was mostly calm with some periods of speeds up to 10
m/s. During one day there was full sunshine. The measured RMS error is 
very constant with a peak to peak variation of less than 2$\mu$m on an
average of 14$\mu$m. The differences are fully consistent with the
allowed errors under environmental changes and also of the same order
as the measurement accuracy. We believe that the significantly better
overall result is mainly due to the much drier and stabler atmosphere
during these measurements as compared with the summer data from the
VertexRSI antenna.

\section{Conclusions}
\label{conclusions}

We have successfully performed a holographic measurement and
consecutive panel setting of the reflectors of the two ALMA prototype
antennas to an accuracy of better than 20 $\mu$m.  Our estimated
measurement accuracy is approximately 5 $\mu $m. The data collection
and analysis software packages are easy to use and provide quick
results of the measurements, directly usable for a panel adjustment
setting. We consider this system suitable for the routine setting of
the ALMA production antennas to the goal of 20 $\mu$m accuracy in an
acceptable time span. Modern survey equipment enables contractors to
deliver reflectors with an accuracy of 50-60 $\mu$m without undue
cost. Although the holography system can easily start with a much
larger error, in the former case it is feasible to reach the
specification with only one panel setting based on holography. We note
that these measurements, being performed at one elevation angle only,
do not provide information on the gravitationally induced deformation
as function of elevation angle.

In summary:

\begin{enumerate}
\item The holography system has functioned according to specification
  and has enabled us to measure the surface of the antenna reflector
  with a repeatability of better than $10\mu$m.
\item As shown in Figs.~\ref{fig:holosVertex} and
  \ref{fig:holosAEC}, we have set both antenna surfaces to and
  accuracy of 16-17 $\mu$m RMS. This will provide an aperture
  efficiency of about 65 percent of that of a perfect reflector at the
  highest observing frequency of 950 GHz.
\item The small differences in the surface maps obtained over several
  days of measurement are consistent with the measurement
  repeatability and at best marginally significant. If taken at face
  value, they indicate that the deformations of the reflector under
  varying wind and temperature influence are fully consistent with,
  and probably well within, the specification.  This excellent
  behaviour over time is more important than the actual achieved
  surface setting. We stopped iteration of the settings after having
  achieved the goal of less than 20 $\mu$m.
\item Further information on the performance of the ALMA Prototype
  Antennas can be found in \cite{Mangum2006}.
\end{enumerate}

\appendices

\section{Holography Map Parameter Equations and Calculations}
\label{equationsandcalcs}

In this appendix we list the equations used to derive the holography
measurement values listed in Table \ref{tab:holoparms}.  The
calculation leading to the power-related expressions of this appendix
are detailed in \cite{Daddario1982}.

\begin{eqnarray}
f_1 &\equiv& \textrm{taper factor for signal feed} \nonumber \\
    &=& 1.13 (6 + 2.5~\textrm{dB taper}) \\
f_{apo} &\equiv& \textrm{apodization smoothing factor} \\
f_{osr} &\equiv& \textrm{map oversampling factor between rows} \\
f_{oss} &\equiv& \textrm{map oversampling factor along a row} \\
D &\equiv& \textrm{main antenna diameter} \\
d & \equiv& \textrm{reference feed diameter} \\
\theta_{ext} &\equiv& \textrm{angular extent of map (assumed square)} \\
\theta_b &\equiv& \textrm{primary beam size} \nonumber \\
         &=& \frac{f_1 c}{\nu D} \nonumber \\
         &=& \frac{61836.6 f_1}{\nu(GHz)D(m)}~\textrm{arcsec} \\
\theta_{sr} &\equiv& \textrm{sampling interval between rows} \nonumber \\
           &=& \frac{\theta_b}{f_{osr}} \nonumber \\
           &=& \frac{f_1 c}{f_{osr}\nu D} \nonumber \\
           &=& \frac{61836.6 f_1}{f_{osr} \nu(GHz)
             D(m)}~\textrm{arcsec} \\
\theta_{ss} &\equiv& \textrm{sampling interval along a scan} \nonumber\\
           &=& \frac{\theta_b}{f_{oss}} \nonumber \\
           &=& \dot{\theta} t_{samp} \nonumber \\
           &=& 0.012 \dot{\theta}~\textrm{arcsec}
\end{eqnarray}

\begin{eqnarray}
N_{row} &\equiv& \textrm{number of rows in map} \\
\delta_d &\equiv& \textrm{spatial resolution on dish} \nonumber \\
         &=& \frac{D}{N_{row}} \nonumber \\
         &=& \frac{f_1 f_{apo} c}{\nu \theta_{ext}} \nonumber \\
         &=& \frac{1717.7 f_1 f_{apo}}{\nu(GHz)
  \theta_{ext}(deg)}~\textrm{cm} \\
\dot{\theta} &\equiv& \textrm{map row scanning rate} \\
L_m &\equiv& \textrm{linear size of map} \\
P &\equiv& \textrm{Transmitter EIRP} \\
P_r &\equiv& \textrm{Reference feed power received} \nonumber \\
    &=& \frac{\pi d^2}{4}\frac{P}{4\pi R^2} \nonumber \\
    &=& \frac{1}{16}\left(\frac{d}{R}\right)^2 P \\
P_s &\equiv& \textrm{Main antenna power received on boresight} \nonumber \\
    &=& \frac{\pi D^2}{4}\frac{P}{4\pi R^2} \nonumber \\
    &=& \frac{1}{16}\left(\frac{D}{R}\right)^2 P \\
P_s\left(\alpha\right) &=& P_s\left(0\right)
\left[\frac{J_1\left(\frac{\pi\alpha
      D}{\lambda}\right)}{\left(\frac{\pi\alpha
      D}{2\lambda}\right)}\right]^2 \\
\sigma^2 &=& \frac{\left[kT_{sys}B + P_r +
P_s\left(\alpha\right)\right]kT_{sys}}{t_{int}} \\
\delta z &=&
         \frac{\lambda}{16\sqrt{2}}\frac{\sqrt{N_{sx}N_{sy}}}{f_{os}^2}
         \frac{\sigma_{av}}{M_0} \nonumber \\
         &=& 0.044\lambda\frac{\sqrt{N_{sx}N_{sy}}}{f_{os}^2}
         \frac{\sigma_{av}}{M_0} \\
B &\equiv& \textrm{Detector bandwidth} \\
t_{int} &\equiv& \textrm{Integration time} \nonumber \\
       &=& N_{row} t_{row} \nonumber \\
       &=& \frac{f_{osr} \theta^2_{ext}}{\dot{\theta} \theta_b} \nonumber \\
       &=& \frac{1717.7\times10^2 f_{osr} f_1 D(m)}
                 {\dot{\theta}(arcsec/sec) \nu(GHz)
                   \delta^2_d(cm)}~\textrm{hours} \\
\alpha &\equiv& \textrm{Scan angle, which ranges over}
                \pm\frac{\theta_{ext}}{2} \\
R &\equiv& \textrm{Distance between holography Tx and Rx} \\
\Delta z &\equiv& \textrm{Reflector surface displacement accuracy}
\end{eqnarray}

For the ALMA holography system:

\begin{enumerate}
\item $t_{int}$ = 36 msec,
\item $M_0$ = $\sqrt{P_s(0) P_r}$ = $4.167\times
10^{-7}$P,
\item $\sigma_0$ = $\left(1.23\times10^{-22}W (P)\right)^\frac{1}{2}$,
\item $P_r$ Term = $\left(2.13\times10^{-27}W (P)\right)^\frac{1}{2}$
\item Average map noise for complex correlator ($\sigma_{av}$) = $\left(2.23\times10^{-25}W (P)\right)^\frac{1}{2}$,
\item $\delta z$ = $\frac{1.35\times 10^{-2}}{\sqrt{P}}$
\end{enumerate}

Thus, if we want an error in the measurement of the surface shape of
$\delta z$ = 5 $\mu$m, we need a transmitter with an EIRP of P = 7.3
$\mu$W. The expected radiated power is in excess of 10$\mu$W, so there
is a good margin. Noise will not be the limiting factor in the accuracy of the
measurement.

\section{Surface RMS Calculation Details}
\label{rmscalc}

The RMS in the holography plots is computed in the following way:

\begin{itemize}

\item {\em Unweighted}:

\begin{equation}
\sigma_u = \sqrt{\frac{1}{N} \sum_i d_i^2 - \left(\frac{1}{N}
    \sum_i d_i\right)^2} 
\label{eq:sigmau}
\end{equation}

\begin{itemize}
\item The summation is on the $N$ unmasked pixels 
\item $d_i$ is the normal surface displacement for pixel $i$
\end{itemize}

\item {\em Weighted}:

\begin{equation}
\sigma_w = 
\sqrt{ \frac{1}{\sum_i w_i} \sum_i w_i e_i^2 - \left(\frac{1}{\sum_i
      w_i} \sum_i w_i e_i\right)^2}
\label{eq:sigmaw}
\end{equation}

\begin{itemize}
\item The summation is on the unmasked pixels
\item $e_i$ is half of the path-length error due to the surface
    displacement for pixel $i$. This is directly related to the
    observed phase errors $ \delta \phi_i$ by:

\begin{eqnarray}
e_i &=& \delta \phi_i \frac{\lambda}{4 \pi} \nonumber \\
    &=& d_i \cos{\alpha_i}
\label{eq:ei}
\end{eqnarray}

\item $w_i$ is the illumination amplitude of an ideal ALMA receiver
    at pixel $i$. This is currently specified to a 12dB taper ($w_i$
    is 0.251 at the edge of the dish). The function used is parabolic:

\begin{equation}
w_i = 1-\left(1-10^{-0.6}\right)\left(\frac{r_i}{6}\right)^2
\label{eq:wi}
\end{equation}

\item $\cos(\alpha)$ is a projection factor which attenuates the 
  effect on antenna efficiency of surface errors for rays close to the
  edge of the dish. This projection factor is:

\begin{equation}
\cos{\alpha_i} = \frac{1}{\sqrt{ 1. + \frac{r^2_i}{4 f^2}}}
\label{eq:cosalpha}
\end{equation}

  \end{itemize}

\end{itemize}

\section*{Acknowledgements}
We gratefully acknowledge the support by the staff of NRAO. In
particular, we thank Nicholas Emerson, Fritz Stauffer, Jack Meadows
and Angel Ot\'arola for their work on the soft- and hardware and panel
adjustments.


\bibliographystyle{IEEEtran}
\bibliography{IEEEabrv,HoloNearFieldPaperIEEEbib}
%

%

\begin{biography}[{\includegraphics[width=1in,height=1.25in,clip,keepaspectratio]{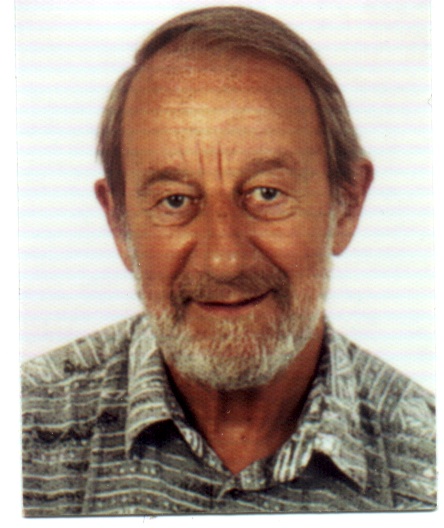}}]{Jacob W.~M.~Baars} received the M.Sc. and
  D.Sc. degrees in applied physics from the Technical University of
  Delft, the Netherlands in 1963 and 1970, respectively. 

After working at the National Radio
Astronomy Observatory in Green Bank, WV, he joined the Netherlands
Foundation for Radio Astronomy in 1969. There he participated in the
construction and was later head of the Westerbork Synthesis Radio
Telescope. In 1975 he joined the Max-Planck-Institut f\"ur
Radioastronomie in Bonn as head of the Division for Millimeter
Technology. He was project Manager of the 30-m Millimeter Radio
Telescope of IRAM in Spain and the Heinrich Hertz Telescope in
Arizona. From 1997-99 he worked on the UMass-Mexico Large Millimeter
Telescope. In 1999 he joined the European Southern Observatory, where
he was involved in several aspects of the ALMA Project, lastly the
evaluation of the prototype antennas. Since his retirement he consults
in the area of large antennas and radio telescopes. His research
interests are in the area of antenna theory and practice, in
particular the performance calibration of large antennas with radio
sources, and of atmospheric influences on observations at very high
frequencies. 

Max-Planck-Institut f\"ur Radioastronomie, Auf dem H\"ugel 69, D-53121
Bonn, Germany and European Southern Observatory, D-85748 Garching,
Germany (jacobbaars@arcor.de)
\end{biography}

\begin{biography}[{\includegraphics[width=1in,height=1.25in,clip,keepaspectratio]{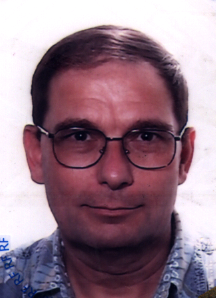}}]{Robert Lucas}
Robert Lucas received a Doctorat d'Etat from Paris University in 1976. After
working with the Meudon Observatory and the Ecole Normale Sup\'erieure in
Paris, he joined the staff of the Grenoble University then the Institut de
RadioAstronomie Millm\'etrique (IRAM) in Grenoble in 1987. He was closely
associated with the commissioning of IRAM telescopes. He is now in charge of
science software requirements for the ALMA project. His research
interests include: line formation processes and chemistry in molecular clouds
and circumstellar envelopes of evolved stars, the characterization of
reflector antennas, the calibration of millimeter-wave interferometric data,
and software for radioastronomy observations.

Institut de Radio Astronomie Millim\'etrique, 300 rue de la Piscine,
Domaine Universitaire, 38406 Saint Martin d'H\'eres, France (lucas@iram.fr)
\end{biography}

\begin{biography}[{\includegraphics[width=1in,height=1.25in,clip,keepaspectratio]{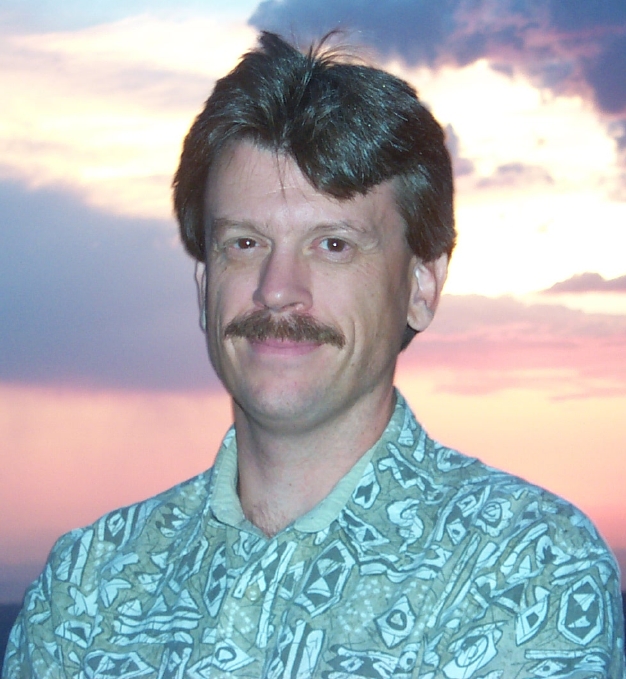}}]{Jeffrey G.~Mangum}
received the Ph.D. degree in astronomy from the University of Virginia
in 1990.  

Following a two-year residency as a postdoctoral researcher in the
astronomy department at the University of Texas he joined the staff of
the Submillimeter Telescope Observatory (SMTO) at the University of
Arizona.  In 1995 he joined the scientific staff at the National Radio
Astronomy Observatory (NRAO) in Tucson, Arizona, and subsequently moved to
the NRAO headquarters in Charlottesville, Virginia.  His research
interests include the astrophysics of star formation, the solar
system, and external galaxies, the performance characterization of
reflector antennas, and calibration of millimeter-wavelength
astronomical measurements.

National Radio Astronomy Observatory, 520
Edgemont Road, Charlottesville, VA  22903, USA (jmangum@nrao.edu)
\end{biography}

\begin{biography}[{\includegraphics[width=1in,height=1.25in,clip,keepaspectratio]{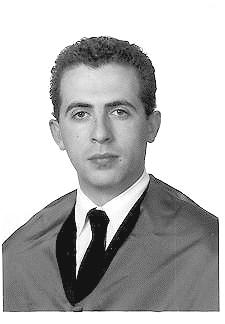}}]{J.~A.~Lopez-Perez}
was born in Martos (Jaen), Spain, in 1971. He received the
Telecommunication Engineer degree from the Universidad Politecnica de
Madrid (UPM) in 1996.

Following a short period of time working at the Radiation Department of 
the UPM, he joined the Institut de Radio Astronomie Millimetrique 
(IRAM) in Grenoble (France) for two years. There he worked in the field 
of microwave antenna holography with David Morris and participated 
actively in the IRAM 30 meter radiotelescope holography campaigns during 
1998, 1999 and 2000. In addition, he worked in the Backend Group of IRAM 
in the design and construction of electronic circuits for the Plateau de 
Bure Interferometer correlator.

After that, he joined the staff of the Observatorio Astronomico Nacional 
(OAN) in Yebes (Spain). There he was involved in the construction of the 
OAN 40 meter radiotelescope. He was also responsible for the design and 
construction of the 40-m holography receiver system and he led the 
construction of the digital autocorrelator spectrometer based in IRAM 
correlator boards.

In 2002, he worked in a feasibility study of a holography measurement 
system for the European Space Agency (ESA) ground antennas.

In 2003, he joined the ALMA Antenna Evaluation Working Group (AEWG) and 
worked in the holographic surface evaluation of the ALMA antenna 
prototypes in Socorro (NM,USA).

He is currently collaborating with INDRA company and the European Space 
Operation Center (ESOC) in the development of a holography system for 
the ESA Deep Space Antenna (DSA) system. He is also leading the construction 
of the 40-m X-band receiver for Very Long Baseline Interferometry (VLBI).

His main research interests are microwave holography of large reflector 
antennas and radioastronomy receiver design and development.

Centro Astronomico de Yebes, Apartado 148, E-19080
Guadalajara, Spain (ja.lperez@oan.es).
\end{biography}





\end{document}